\documentclass[twocolumn,prb]{revtex4}
\usepackage{graphicx}

\def\prb{Phys. Rev. B }
\def\prl{Phys. Rev. Lett. }

\def\be{\begin{equation}}
\def\ee{\end{equation}}
\def\ba{\begin{eqnarray}}
\def\ea{\end{eqnarray}}

\begin{document}

\title{Phase diagram and isotope effects of the quasi-one-dimensional electron gas \\ coupled to
phonons}

\author{Ian P. Bindloss}
\affiliation{Department of Physics, University of California,  
Los Angeles, California 90095-1547}
\date{\today}

\begin{abstract}
Using a multistep renormalization group method, we study the low-temperature phases of the
interacting one-dimensional (1D) electron gas coupled to phonons.  We obtain analytic expressions
for the weak-coupling quantum phase boundaries of the 1D extended Holstein-Hubbard model and the
1D extended Peierls-Hubbard model for general band-filling and phonon frequency.  Away from
half-filling, the phase diagrams are characterized by a delicate competition between spin density
wave, charge density wave, and superconducting orders.  We study the dependence of the ground
state on the electron-phonon (el-ph) and electron-electron (el-el) coupling strengths, the
screening length, electron bandwidth, phonon frequency, doping, and type of phonon.  Unlike the
case in Fermi liquids, in 1D the el-ph coupling is strongly renormalized, often to stronger
values.  Even when the bare phonon-induced attraction is weak compared to the bare el-el
repulsion, a small amount of retardation can cause the renormalized el-ph interaction to dominate
the problem.  We find cases in which a repulsive el-el interaction enhances the superconducting
susceptibility in the presence of a retarded el-ph interaction.  The spin gap and superconducting
susceptibility are found to be strongly dependent on the deviation from half-filling (doping). 
In some cases, the superconducting susceptibility varies nonmonotonically with doping and
exhibits a maximum at a particular doping.  For a quasi-1D array of weakly coupled, fluctuating
1D chains, the superconducting transition temperature $T_c$ also exhibits a maximum as a function
of doping.  The effect of changing the ion mass (isotope effect) on $T_c$ is found to be largest
near half-filling and to decrease rapidly with doping.
\end{abstract}

\maketitle

\section{Introduction}
\label{introduction}
 
Recent experiments\cite{shen1,gweon} in the high-temperature superconductors that
suggest a strong and ubiquitous
electron-phonon (el-ph) interaction have lead to an increased interest in strongly correlated
el-ph systems.
Of particular interest is the interplay between the
repulsive, instantaneous Coulomb interaction and the attractive, retarded interaction mediated by
phonons, in the context of various phase transitions (instabilities) of the system.
This interplay is quite simple and well understood in a two- or three-dimensional Fermi liquid,
and serves
as a foundation for the BCS theory of superconductivity.  However, the high-temperature 
superconductors,
as well as some other materials of current interest, exhibit manifestly non-Fermi liquid
behavior.\cite{anderson,vallaNFL,orenstein,drorprl,ian}

Compared to conventional metals,
much less is known about the influence of an el-ph interaction
in non-Fermi liquids. 
One such system, where much analytic progress can be
made,\cite{Fradkin,Fradkin2,Steve,Voit1,Voit2}
is the interacting
one-dimensional electron gas (1DEG), where the system is strongly
correlated even for weak interactions.
In contrast to higher dimensions, in one-dimension
the el-ph interaction
is strongly renormalized.  In other words,
the effective el-ph interaction at low energies is not
simply given by the bare, microscopic coupling constants,
which makes the physics both very
different and very rich.  Furthermore, the renormalizations of the el-ph interactions are
affected by
direct electron-electron (el-el) interactions.
It is unknown how much of the physics of the 1DEG is generic to
other non-Fermi liquids; but one may hope that certain features are generic, in which case the
1DEG may serve
as a paradigmatic model for a broader class of systems. 
In addition, there are many reasons to study the 1DEG coupled to phonons for its own sake, 
including the importance of el-ph interactions in the quasi-1D conducting
polymers,\cite{KivelsonRMP,gruner}
and the possibility that they play an important role in the quasi-1D organic
conductors.\cite{bourbonnais}
It may also be the case that holes in the high-temperature superconductors live in quasi-1D due
to 
stripe correlations.\cite{stripereview}
In any case, since in 1D a small bare el-ph coupling can be renormalized to
substantially larger values, the el-ph interaction is important to consider.

In the present paper, we use a multistep renormalization group (RG) procedure to
comprehensively study the zero-temperature phase
diagram of the spinful 1DEG coupled to phonons, treating the el-el
and el-ph interactions on equal footing.  The same technique is employed to compute the doping
dependent superconducting susceptibilities, charge density wave susceptibility, and isotope
effects.
In a separate paper,\cite{ian} we have studied
the influence of the el-ph interaction on the
electron dynamics of an interacting 1DEG, expressed via the single particle spectral function.
The strategy in the present paper is to start with a microscopic electron-phonon model with many
parameters
(el-el interactions, electron bandwidth, el-ph interactions, and phonon frequency),
and then to integrate out high-energy degrees of freedom to produce
a low-energy effective field theory with a known phase diagram--the continuum 1DEG--whose only
parameters are the renormalized el-el interactions and bandwidth.

If the system is far from commensurate filling, we employ the ``two-step RG''
procedure.\cite{Grest,Caron,Steve}  In this method,
one set of RG equations governs the flow
of the coupling constants for energies $\omega$ greater than the phonon frequency $\omega_0$,
while a second set governs the flow for $\omega_0 > \omega$.  If, as usual, the
Fermi energy $E_F > \omega_0$, the first step is to integrate out degrees of freedom
from $E_F$ to $\omega_0$ using the microscopic coupling constants as initial values.
The resulting renormalized couplings are then used as initial values
in the second stage of RG flows, to integrate out degrees of freedom from $\omega_0$
to some low-energy scale.  We also use the two-step RG technique to study systems at half-filling
(the only difference being the inclusion of Umklapp scattering). 
Near, but not equal to half-filling, there are three steps to the RG
transformation.  Defining $\mu$ as the chemical potential relative to its
value at half-filling, for $\mu > \omega_0$ these steps are
$E_F > \omega > \mu$, then $\mu > \omega > \omega_0$, and finally $\omega_0 > \omega$.  This
``three-step RG'' technique allows us to study the continuous evolution of the phase diagram with
doping.

We study the continuum limit of two microscopic
models of interacting, spinful 1D electrons coupled to phonons: the extended Holstein-Hubbard
model\cite{Holstein} and extended Peierls-Hubbard model.  While others have employed RG
techniques to study the 1DEG coupled to phonons,\cite{Grest,Caron,Steve,Voit1,Voit2} and indeed
some of the qualitative results of the present paper have been known for some time,
the aforementioned models have not been explored in detail with multistep RG.  Numerical
calculations of their phase diagrams have been mostly limited to simplified models that contain
spinless electrons, infinite ion mass, or zero el-el interactions.\cite{Bursill,Weisse}  These
studies have also been mostly limited to specific values of the band filling, especially
half-filling.\cite{Sengupta}

We obtain analytic expressions for the low-temperature phase boundaries of
the aforementioned models.
Since the RG procedure is perturbative (one-loop), our results are
only accurate for interactions that are small compared to the bandwidth,
but are valid for any relative strength of the el-ph and el-el interactions.
Corrections at stronger
couplings are expected to be smooth and should not make large qualitative changes to the results,
as long as the interactions are not too large.
The method properly takes into account the quantum phonon dynamics,
and is therefore used to study phonons of nonzero frequency.
 	
One question we address is whether superconductivity can exist in
realistic quasi-1D systems in which the bare
el-el repulsion is stronger than the bare attractive interaction mediated by
the el-ph coupling.  It is known that in the absence of el-ph interactions, a repulsive el-el
interaction in a single-chain 1DEG is always harmful to superconductivity, as is the case in
higher dimensions.  However, we find
that for the 1DEG coupled to phonons, 
increasing the el-el repulsion can in some specific cases enhance superconductivity.
Moreover, with even a small amount of retardation present,
it is possible for a 1DEG to have a divergent superconducting
susceptibility even when the bare el-el repulsive is much stronger than the bare el-ph
attraction.
(In three dimensions, this is only possible with substantial retardation.)

In ordinary metals, the observation of an isotope effect on $T_c$
was crucial in the development of the
BCS theory that describes the superconductivity of Fermi liquids. 
The isotope effect exponents
$\alpha_{T_c} = - d \ln T_c / d \ln M$ 
and $\alpha_{\Delta} = - d \ln \Delta / d \ln M$
have the universal value of $1/2$, where $M$ is the ion mass and $\Delta$
is the superconducting gap.

In contrast,
in the cuprate high-temperature superconductors, both $T_c$ and $\alpha_{T_c}$
are strongly doping-dependent.
Despite the fact that the superconducting gap
is a monotonically decreasing function of increasing doping,
$T_c$ varies nonmonotonically with doping, exhibiting a maximum at ``optimal doping.''
For dopings well below optimal, the isotope effect on $T_c$ is quite large: $\alpha_{T_c}\approx
1$.
As the doping increases, $\alpha_{T_c}$ decreases, usually
dropping below 0.1 near optimal doping.\cite{isotope}  The isotope effect on the so-called
pseudogap
has the opposite sign as the isotope effect on $T_c$.\cite{pseudogap}
The origin of
these highly unconventional isotope effects remains one of the many
unsolved mysteries of high-temperature superconductivity.  

Since the 1DEG is perhaps the only presently solvable non-Fermi liquid, it is worth computing
isotope effects in this system to try to gain insights on isotope effects in unconventional
superconductors.
We compute $\alpha_{T_c}$ for a quasi-1DEG coupled to phonons, under the
assumption that charge density wave order is dephased by spatial
or dynamic fluctuations of the 1D chains.\cite{zachar,Nature}
For most choices of the parameters, $\alpha_{T_c}$ is larger than the BCS value
at small dopings, then drops below $1/2$ as the doping is increased.
We show that the quasi-1DEG coupled to phonons displays a
strongly doping-dependent
$T_c$ that can exhibit a maximum as a function of doping.
This behavior occurs despite the fact that the pairing energy, determined by the spin gap
$\Delta_s$,
is a monotonically decreasing function of increasing doping.
We also compute the isotope exponent $\alpha_{\Delta_s} = - d \ln \Delta_s / d \ln M$ and find 
$\alpha_{\Delta_s} < 0$, which in most cases is the opposite sign as $\alpha_{T_c}$.

The rest of this paper is organized as follows.
Section \ref{models} defines the microscopic models.
Section \ref{results} presents our results for the phase diagrams, without derivation.
In Section \ref{isotopesection} we give our results for
the doping dependence of the superconducting
susceptibility and isotope effects, again without derivation.
In Section \ref{compare} we compare our analytical results for the phase diagrams to
numerical work of other authors.  Section \ref{technique}
discusses the RG flows of the coupling constants and
contains a mathematical derivation of all the results in the previous sections.
In Section \ref{conclusion} we summarize the results qualitatively and
make some concluding remarks.

\section{Models of 1D electron-phonon systems}
\label{models}

The 1D extended Peierls-Hubbard (Pei-Hub) model is
defined by the Hamiltonian
\ba
\label{Pei}
\nonumber {\cal H}_{\rm Pei-Hub} &=& -t \sum_{i,\sigma}[1 - \tilde \gamma(a_i^\dagger +
a_i)](c_{i,\sigma}^\dagger c_{i+1,\sigma} + {\rm H.c.})\\
&+& \omega_0 \sum_i a_i^\dagger a_i \; +  \; {\cal H}_{UV},
\ea
where the el-el interaction part is that of the extended Hubbard model:
\be
{\cal H}_{UV} = U \sum_i n_{i,\uparrow} n_{i,\downarrow} + V \sum_i n_i n_{i+1}  .
\ee
Here, $c_{i,\sigma}^\dagger$ creates an electron of
spin $\sigma$ on site $i$, $a_i^\dagger$ creates a phonon of frequency $\omega_0$ between sites
$i$ and $i+1$, $n_i = \sum_{\sigma} n_{i,\sigma} = \sum_{\sigma} c_{i,\sigma}^\dagger
c_{i,\sigma}$, and $\tilde{\gamma}$ is the dimensionless el-ph coupling constant.  This model,
in the absence of ${\cal H}_{UV}$, is an approximation to the model of Su, Schrieffer, and
Heeger\cite{SSH} (SSH).  Including extended Hubbard interactions, the SSH model is
\ba
&& \nonumber {\cal H}_{\rm SSH-Hub} = \\
&& \nonumber -\sum_{i,\sigma}\left[t - \gamma(u_{i+1}-u_{i})\right](c_{i,\sigma}^\dagger
c_{i+1,\sigma} + {\rm H.c.})\\
&& + \, \sum_i \left[ \frac{p_i^2}{2M} + \frac{\kappa}{2} (u_{i+1}-u_i)^2 \right] +  {\cal
H}_{UV}. \;\;\;\;
\label{SSHeq}
\ea
Here, acoustic phonons with spring constant $\kappa$ couple to electrons by modifying the bare
hopping matrix element $t$ by the el-ph coupling strength $\gamma$ times the relative
displacements $u_{i+1} - u_i$ of two neighboring ions of mass $M$.
If we approximate the acoustic phonon as an Einstein phonon of frequency $2 \sqrt{\kappa/M}
\equiv \omega_0$, the SSH-Hub model reduces to the Pei-Hub model.  This is a good approximation
since, in the SSH model, the el-ph interaction vanishes at zero momentum transfer.  The el-ph
coupling constants of the models are related via $\gamma = \tilde \gamma \, t\sqrt{2 M
\omega_0}$.

The 1D extended Holstein-Hubbard (Hol-Hub) model
is defined by the Hamiltonian
\ba
&&\nonumber {\cal H}_{\rm Hol-Hub} = \\
&& \nonumber -t\sum_{i,\sigma}(c_{i,\sigma}^\dagger c_{i+1,\sigma} + {\rm H.c.}) +  \sum_i \left[
\frac{p_i^2}{2M} + \frac{1}{2}M \omega_0^2 q_i^2 \right]  \\
&& + \: g \sqrt{2 M \omega_0} \, \sum_i q_i n_i + {\cal H}_{UV}  .
\label{Hol}
\ea
In this model, a dispersionless optical
phonon mode with vibrational coordinate $q_i$ and frequency
$\omega_0$
couples to the local
electron density with el-ph coupling strength $g$.
Unlike the Pei-Hub model, this model contains equal parts 
backward scattering (momentum transfer near $2 k_F$)
{\it and} forward scattering
(momentum transfer near 0) el-ph interactions.
Another difference is that
the el-ph interaction is site centered (diagonal) in the
Hol-Hub model versus bond centered (off-diagonal) in the Pei-Hub model. 
In the present paper,
we only explicitly discuss the case of repulsive el-el interactions
($U,\, V \ge 0)$; however, the mathematics remains valid for attractive ones. 

For convenience we
define the following dimensionless quantities:
\ba
\label{couplingconstants}
\nonumber&& \lambda_{\rm Pei} = \frac{4 \gamma^2 \sin^2 k_F}{\pi v_F \kappa} , \;\;\; 
\lambda_{\rm Hol}  = \frac{2 g^2}{\pi v_F \omega_0}  ,\\
&& {\bar U} = \frac{U}{\pi v_F}  , \;\;\; {\bar V} = \frac{V^\prime}{\pi v_F} , \;\;\; l_0 =
\ln\left(\frac{E_F}{\omega_0}\right) ,\;\;\;\; 
\ea
where
$v_F = 2t \sin(k_F)$,
$V^\prime = -V \cos(2 k_F)$,
$E_F > \omega_0$ is a high-energy cutoff for the RG theory
on the order of the Fermi energy, and $2 k_F/\pi$ is the average number
of fermions per site.
The el-ph coupling parameters $\lambda_{\rm Pei}$
and $\lambda_{\rm Hol}$
are defined such that, in the absence of el-el interactions, the spin gap
is given by $\Delta_s \propto \exp(-a/\lambda)$, where $a = 1/2$ for half-filling,
$a = 1$ for incommensurate fillings, and $\lambda$ stands for $\lambda_{\rm Pei}$ or
$\lambda_{\rm Hol}$, depending on the model.  (In Section \ref{technique} we give
the result for $\Delta_s$ is the presence of el-el interactions.)
The method we employ yields phase diagrams that
are accurate for $\lambda, \,  {\bar U}, \, {\bar V} \ll 1$.  We expect that
the technique is qualitatively accurate when these couplings are of order 1.
It should be clear that, in the present paper,
we have set the lattice parameter and Planck's constant equal to 1.

\section{Results for the phase diagrams}
\label{results}

In this section, we present our main results for the phase diagrams, without derivation.  
More discussion of the method and a detailed derivation are given in Section \ref{technique}, where
we obtain explicit expressions for the phase boundaries. 

\subsection{Incommensurate filling}

Below we present zero-temperature phase diagrams in the incommensurate limit,
which corresponds to $\mu \sim E_F$.

\subsubsection{Transition to the spin-gapped phase}

For the 1D extended Hubbard model without el-ph interactions,
the low-energy properties are described by the spin-charge separated Luttinger liquid (LL)
as long as ${\bar V} < {\bar U}/2$.
The low energy properties of this gapless, quantum-critical state of matter 
can be described by a bosonic free field theory.\cite{emery,giamarchi}
Since the quasiparticle residue vanishes, the LL is by definition
a non-Fermi liquid; there are no elementary excitations with the quantum
number of an electron (or a hole).

In the absence of el-ph interactions,
a spectral gap develops in the spin sector if ${\bar V} > {\bar U}/2$,
which leads to quite different
physical properties than the LL.
This non-Fermi liquid phase is termed a Luther-Emery liquid\cite{LEL} (LEL).
However, in nature one typically expects ${\bar V} < {\bar U}/2$,
so that some additional physics is needed to created a spin gap.
For incommensurate fillings, the charge sector is gapless.

\begin{figure}
\includegraphics[width=0.36\textwidth]{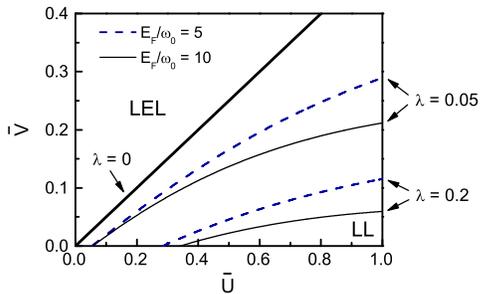}
\caption{\label{bindloss_fig_1} Phase diagram in the ${\bar V}$-${\bar U}$ plane showing the
$\lambda$
dependence of the phase boundary between the gapless
Luttinger liquid (LL) phase and the spin-gapped Luther-Emery liquid (LEL) phase, for an
incommensurate 1DEG with $E_F/\omega_0 = 5$ (dashed lines) and $E_F/\omega_0 = 10$ (thin solid
lines).
The thick line shows the
phase boundary at $\lambda = 0$ for any $E_F/\omega_0$.
$\lambda$ stands for either $\lambda_{\rm Hol}$ or $\lambda_{\rm Pei}$, depending on the model.
}
\end{figure}

\begin{figure}
\includegraphics[width=0.3\textwidth]{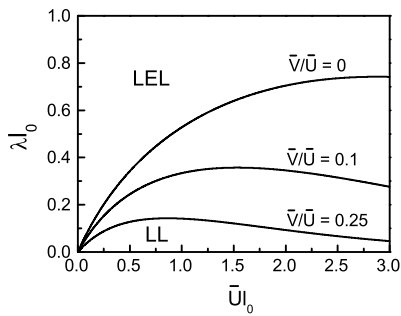}
\caption{\label{bindloss_fig_2} Phase diagram in the $\lambda l_0$-${\bar U}l_0$ 
plane for an incommensurate 1DEG,
showing the dependence on ${\bar V}/{\bar U}$ of the phase boundary separating the LL and LEL
phases.
$\lambda$ stands for either $\lambda_{\rm Hol}$ or $\lambda_{\rm Pei}$, depending on the model.
}
\end{figure}

We now study the effects of the el-ph interaction on the phase boundary between the LL and
spin-gapped LEL.
In Fig. \ref{bindloss_fig_1} we show a phase diagram in the ${\bar V}$-${\bar U}$ plane, for
various
fixed values of $\lambda$ and $E_F/\omega_0$ (this phase boundary is the same
for the Pei-Hub and Hol-Hub models).
We see that
a retarded el-ph interaction dramatically increases the stability of the LEL phase relative to
the
LL phase.  The phase boundary is very sensitive to the retardation parameter $E_F/\omega_0$,
with higher values favoring the LEL phase.
For the case of an unretarded el-ph interaction ($\omega_0 > E_F$),
which is not typical in real materials,
a spin gap can only occur for $\lambda > {\bar U} - 2{\bar V}$.
However, just a modest amount of
retardation makes a spin gap possible, even, in some cases, for $\lambda \ll {\bar U} - 2{\bar
V}$.
As we show in Section \ref{technique}, this is
due to the renormalization of the backscattering el-ph interaction toward
stronger values.
The figure shows that poorly screened interactions (high ${\bar V}/{\bar U}$) favor the LEL.
The dependence of the phase boundary on ${\bar V}/{\bar U}$ is shown
explicitly in Fig. \ref{bindloss_fig_2}, which presents a phase
diagram in the $\lambda l_0$-${\bar U}l_0$ plane.  This diagram shows
that scaling eventually carries one to the
case in which an infinitesimal $\lambda$ causes a spin gap (in other words,
for $E_F \gg \omega_0$, the system is spin-gapped for infinitesimal $\lambda$).

\subsubsection{Competition between spin density wave, charge density wave, and superconducting instabilities}
\begin{figure}
\includegraphics[width=0.45\textwidth]{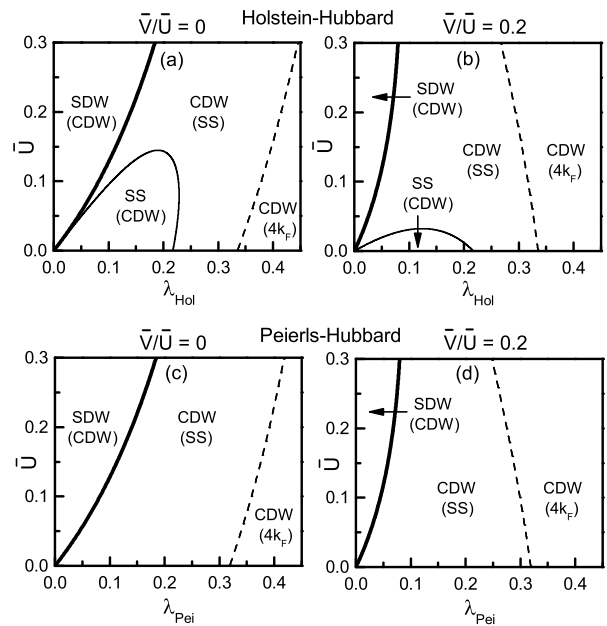}
\caption{\label{bindloss_fig_3} Phase diagrams of the incommensurate extended Hol-Hub model with 
(a) $V = 0$ and (b) ${\bar V}/{\bar U} = 0.2$, and of the incommensurate extended Pei-Hub model
for
(c) $V = 0$ and (d) ${\bar V}/{\bar U} = 0.2$.  For all diagrams, $E_F/\omega_0 = 10$.
To the right of the thick line, the system is spin-gapped.
The most divergent susceptibility is shown without parenthesis,
while parenthesis indicate a susceptibility that diverges less strongly.
SDW stands for $2 k_F$ spin density wave, CDW stands for $2 k_F$ charge density wave,
SS stands for singlet superconductivity, and $4 k_F$ stands for $4 k_F$ charge density wave.}
\end{figure}

\begin{figure}
\includegraphics[width=0.45\textwidth]{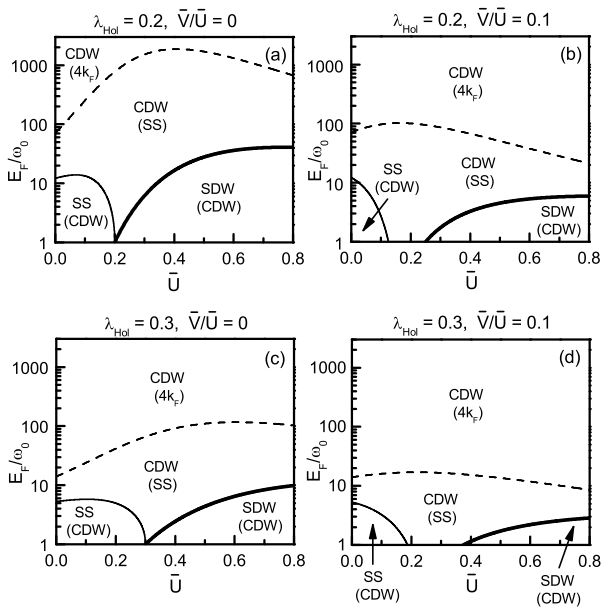}
\caption{\label{bindloss_fig_4} Phase diagrams of the incommensurate extended Hol-Hub model.
(a) and (b) $\lambda_{\rm Hol} = 0.2$; (c) and (d) $\lambda_{\rm Hol} = 0.3$.
(a) and (c) ${\bar V}/{\bar U} = 0$; (b) and (d) ${\bar V}/{\bar U} = 0.1$.  
Parenthesis indicate a subdominant susceptibility.  The region
with dominant SDW is the LL phase, while the rest of the parameter space is a LEL.}
\end{figure}

Below, we explore the many
ordering instabilities present in the system.
$2k_F$ spin density wave order (SDW),
$2k_F$ charge density wave order
(CDW),
and singlet superconductivity (SS) can all compete at zero temperature. 
A divergent charge density wave susceptibility
with $4 k_F$ periodicity (labeled in phase diagrams as ``$4 k_F$'')
is also possible.
It is important to note that
since long range order is forbidden in an incommensurate 1D system,
the phase diagrams below actually consist of identifying
instabilities with divergent response functions.
However, for a quasi-1D array of weakly coupled chains, interchain coupling
allows for true broken symmetry order at low temperature.

In Fig. \ref{bindloss_fig_3} we present phase diagrams in the ${\bar U}$-$\lambda_{\rm Hol}$
and ${\bar U}$-$\lambda_{\rm Pei}$ planes,
for $E_F/\omega_0 = 10$.
In these diagrams, we show
phase boundaries between regions where various ordering fluctuations have divergent
susceptibilities in the low-temperature limit.  The susceptibility
that diverges most strongly, i.e., dominates, is shown without parenthesis.
If a second susceptibility diverges, but less strongly, it is termed ``subdominant,''
and is shown in parenthesis.  The thick solid line
is the LL-LEL transition line; the LL phase 
is present to left of this line and the LEL phase to the right.

For repulsive el-el interactions, the entire LL phase, for either model, is dominated by SDW
fluctuations,
with a slightly weaker CDW susceptibility.
The LEL phase is more complex.  For the extended Hol-Hub model,
dominant SS order is possible provided the el-el repulsion is weak enough
and $\lambda_{\rm Hol}$ is neither too weak {\it nor} too strong.
For the Pei-Hub
model with repulsive el-el interactions, a phase with dominant SS is impossible
due to the absence of el-ph forward scattering.
Therefore, generally speaking, an optical phonon is more favorable
to superconductivity than an acoustic one.
In both
models, there is a large region, for intermediate values of $\lambda$, with
dominant CDW and subdominant SS.
At high values of $\lambda$, SS is no longer divergent;
in this region a $2k_F$ CDW dominates and a $4k_F$ CDW is subdominant. 
We must point out that
the dashed line is not expected to be quantitatively accurate, since the method is
a weak-coupling one. 
Note that the phase with dominant SS is strongly suppressed by poor screening
(large ${\bar V}/{\bar U}$).

\begin{figure}
\includegraphics[width=0.44\textwidth]{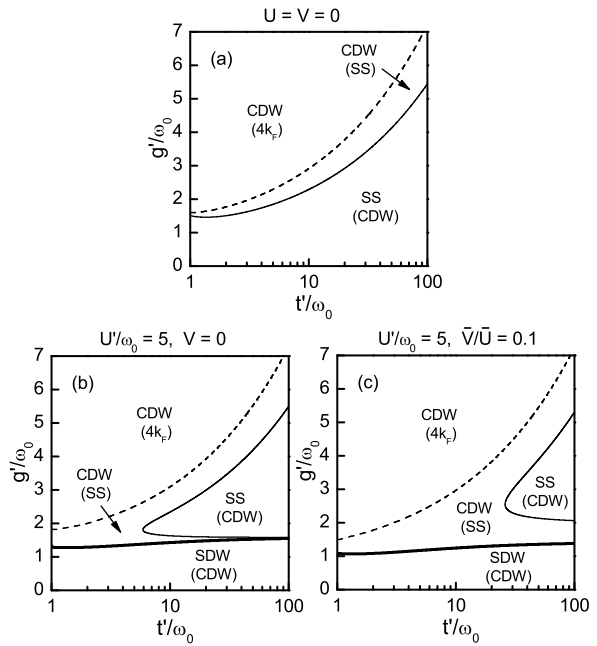}
\caption{\label{bindloss_fig_5} Phase diagram in the $g^\prime/\omega_0$-$t^\prime/\omega_0$
plane of the spinful incommensurate extended Hol-Hub model for (a) $U = V = 0$; (b)
$U^\prime/\omega_0 = 5$, $V = 0$; and (c) $U^\prime/\omega_0 = 5$, ${\bar V}/{\bar U} = 0.1$.
Here $g^\prime \equiv g \sqrt{E_F/v_F}$, $t^\prime \equiv t(E_F/2t)$, and $U^\prime \equiv
U(E_F/v_F)$.
In (a), a spin gap is present
everywhere in the phase diagram, while in (b) and (c), it is present above the thick solid line.
}
\end{figure}  

In Fig. \ref{bindloss_fig_4}, we study the dependence of the ground state on $E_F/\omega_0$,
for the Hol-Hub model,
by showing phase diagrams in the $E_F/\omega_0$-${\bar U}$ plane.
We see that high values of $E_F/\omega_0$ create a spin-gapped phase with dominant CDW.
Low values of $E_F/\omega_0$ create a SS-dominated LEL for low ${\bar U}$,
and a LL for high ${\bar U}$.
In other words, an electron bandwidth that is too big
is harmful to superconductivity!
Note that for moderate $E_F/\omega_0$,
the system lies in the region with dominant CDW and subdominant SS,
which extends from ${\bar U} = 0$ to quite large values of ${\bar U}$.
Therefore, even when the bare interactions are predominantly repulsive
(${\bar U} \gg \lambda_{\rm Hol}$), it is still possible for the system to have a divergent
superconducting correlation.  It is possible for this to occur even with
a small degree of retardation like $E_F/\omega_0 \sim 5$ [see Fig. \ref{bindloss_fig_4}(d)].
The phase diagram of the Pei-Hub model
is similar to Fig.
\ref{bindloss_fig_4}, except that the phase with dominant
SS is removed, and the dashed line is
shifted to slightly lower $E_F/\omega_0$.

It is worth pointing out the intriguing possibility that,
for a {\it quasi}-1D system
with dynamically fluctuating 1D chains,
or even for chains that exhibit transverse spatial
fluctuations, CDW order is easily dephased,
while the superconducting instability is not.\cite{Nature}
If this is the case, then it is possible
for the system to support superconductivity even for the physically realistic
case of ${\bar U} \gg \lambda_{\rm Hol}$, and without the large
amount of retardation that is required in 3D.

A number of authors have computed
the phase diagram of the {\it spinless} Holstein model at half-filling, in
the absence of el-el interactions (see Fig. \ref{bindloss_fig_9}).
To facilitate comparison between this model and the spinful incommensurate extended
Holstein-Hubbard model, Fig. \ref{bindloss_fig_5} shows our phase diagram for the latter model,
in units similar to Fig. \ref{bindloss_fig_9}.

\subsubsection{Enhancement of superconductivity by repulsive interactions}

The intuitive notion that repulsive interactions
suppress superconductivity at weak-coupling, while always true in a Fermi liquid,
does not always hold for the 1DEG coupled to phonons.  
In the 1DEG,
the potentially strongly divergent part of the singlet superconducting susceptibility at
temperature $T \ll \omega_0$ is $\chi_{\rm SS} \propto \Delta_s T^{1/K_c^{\rm eff} - 2}$.
Interactions in the spin channel determine $\Delta_s$, while charge interactions
renormalize the effective Luttinger parameter $K_c^{\rm eff}$ away from its noninteracting value
of 1 (see Section \ref{technique}).
Clearly, SS is enhanced by an increase in $\Delta_s$ or an increase in $K_c$.  In many cases,
increasing the el-el repulsion 
causes both $\Delta_s$ and $K_c$ to decrease.  However, below we discuss cases in which one of
the two parameters is {\it increased} by el-el repulsion; depending on how much the second
parameter is reduced, $\chi_{\rm SS}$ may be enhanced.

In the absence of el-ph interactions, an increase in $U$ always decreases $K_c$ and $\Delta_s$,
and therefore suppresses superconductivity.  In a 1D el-ph system, this is often the case as
well; however, there are also cases in which increasing $U$ can cause $K_c^{\rm eff}$ to be {\it
increased}.  (The technical reason is a decrease in the effective el-ph backscattering
interaction--see Section \ref{technique}.)
Then, if $\Delta_s$ is not reduced too much by the increase in $U$, it is possible for SS to be
enhanced.  An example of this can be seen in Fig. \ref{bindloss_fig_3}(a) or
\ref{bindloss_fig_3}(c).
If we start at $\lambda = 0.35$ and ${\bar U} = 0$, and increase ${\bar U}$ while holding
$\lambda$ fixed, we cross from a region without divergent SS to a region with divergent SS.
This phenomenon can also be seen in Figs. \ref{bindloss_fig_4}(a) and \ref{bindloss_fig_4}(c) if
the right value of $E_F/\omega_0$ is chosen [such as, for example, $E_F/\omega_0 = 200$ in Fig.
\ref{bindloss_fig_4}(a)].
However, if we set ${\bar V}/{\bar U} \ge 1/6$ and hold this ratio fixed while increasing ${\bar
U}$, superconductivity is never enhanced at weak coupling [see Figs. \ref{bindloss_fig_3}(b),
\ref{bindloss_fig_3}(d), and Section \ref{technique}].

\begin{figure}
\includegraphics[width=0.495\textwidth]{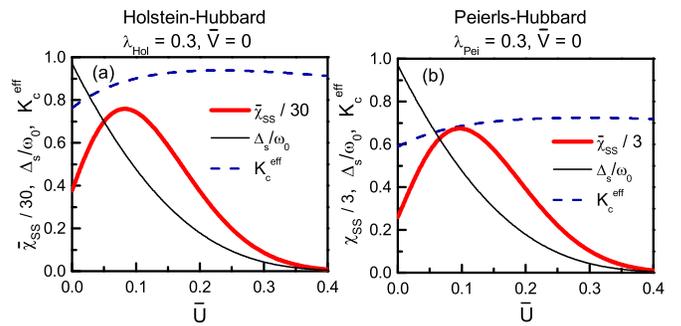}
\caption{\label{increaseU} Dependence of the superconducting susceptibility $\bar \chi_{\rm SS}$
(thick line) on the Hubbard repulsion ${\bar U}$, for an incommensurate system with $V = 0$,
$E_F/\omega_0 = 10$, and $T/\omega_0 = 0.01$.  (a) is for the Hol-Hub model with $\lambda_{\rm
Hol} = 0.3$, and (b) is for the Pei-Hub model with $\lambda_{\rm Pei} = 0.3$.  The thin line is
$\Delta_s/\omega_0$ and the dashed line is the effective Luttinger exponent $K_c^{\rm eff}$.}
\end{figure}

In Fig. \ref{increaseU}, we further investigate the enhancement of 
SS by repulsive interactions by plotting the dimensionless superconducting susceptibility
\be
\label{barchiSS}
\bar \chi_{\rm SS} \equiv \pi v_F \chi_{\rm SS} = (\Delta_s/E_F) (T/E_F)^{1/K_c^{\rm eff} - 2}
\ee
versus $\bar U$, at fixed $\lambda = 0.3$, $V = 0$, and $T/\omega_0 = 0.01$.  For small ${\bar
U}$, $\bar \chi_{\rm SS}$ increases with increasing ${\bar U}$.  However, as ${\bar U}$ is
increased further, 
$K_c^{\rm eff}$ stops increasing as rapidly, and the decreasing $\Delta_s$ causes $\bar \chi_{\rm
SS}$ to drop back down.

\begin{figure}
\includegraphics[width=0.495\textwidth]{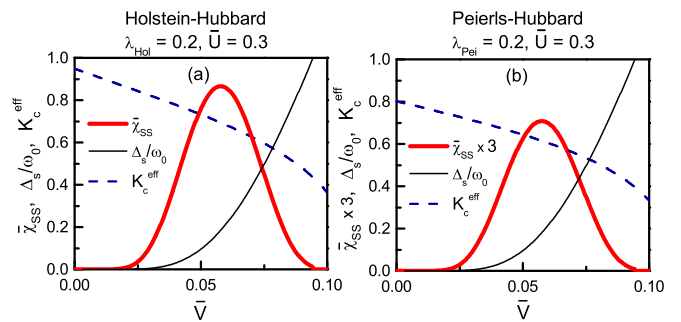}
\caption{\label{increaseV} Dependence of $\bar \chi_{\rm SS}$ (thick line) on the
nearest-neighbor repulsion ${\bar V}$, for an incommensurate system with ${\bar U = 0.3}$,
$E_F/\omega_0 = 10$, and $T/\omega_0 = 0.01$.  (a) is for the Hol-Hub model with $\lambda_{\rm
Hol} = 0.2$, and (b) is for the Pei-Hub model with $\lambda_{\rm Pei} = 0.2$.  The thin line shows
$\Delta_s/\omega_0$ and the dashed line is the effective Luttinger exponent $K_c^{\rm eff}$.}
\end{figure}

It is also possible to enhance superconductivity, in some cases, by increasing the
nearest-neighbor Hubbard repulsion $V$, while holding $U$ fixed.  This causes a renormalization
of 
the el-ph backscattering toward stronger coupling, which results in an increase of $\Delta_s$ and
a decrease of $K_c^{\rm eff}$.  Depending on the competition between these two effects,
$\chi_{\rm SS}$ may (or may not) be enhanced.  An example of a case in which
$\chi_{\rm SS}$ is enhanced can be seen in Figs. \ref{bindloss_fig_3}(a) and
\ref{bindloss_fig_3}(c).  There, if one begins at $\lambda = 0.1$ and ${\bar U} = 0.2$, then
increases ${\bar V}/{\bar U}$ from 0 to 0.2 while holding $\lambda$ and ${\bar U}$ fixed,
the system moves from a LL phase without divergent SS, to a LEL phase with divergent SS.  

In Fig. \ref{increaseV}, we explore this phenomenon further by plotting $\chi_{\rm SS}$ versus
$\bar V$ at fixed ${\bar U} = 0.3$, $\lambda = 0.2$, and $T/\omega_0 = 0.01$.  For small ${\bar
V}$,  $\chi_{\rm SS}$ is enhanced by increasing ${\bar V}$, due to the increase in $\Delta_s$. 
However, as ${\bar V}$ is increased further,
eventually the rapidly decreasing $K_c^{\rm eff}$ begins to overwhelm the effect of the
increasing $\Delta_s$, and $\chi_{\rm SS}$ decreases.  In this case, optimizing superconductivity
therefore requires a fine tuning of ${\bar V}/{\bar U}$.

We have pointed out exceptions to ``rule'' that el-el repulsion suppresses superconductivity at
weak coupling, in order to illustrate, as a point of principle, the dramatically different
physics that governs 1D el-ph systems compared with a Fermi liquid coupled to phonons.  It is
worth briefly discussing some prior works on this topic.  Although the RG flow equations in 
Ref. \onlinecite{Steve} are correct, the authors implied that an enhancement of superconductivity
by repulsive interactions is a generic feature of the 1DEG coupled to phonons,
while Ref. \onlinecite{Voit2} concluded that repulsion always suppresses superconductivity.   Our
results indicate that both works overstated things; indeed, both situations are possible,
depending on the choice of parameters.  The disagreement between
Ref. \onlinecite{Steve} and Ref. \onlinecite{Voit2} was caused, in large part, by the fact that
Ref. \onlinecite{Steve} focused purely on the effect of the el-ph interaction on $\Delta_s$,
while Ref. \onlinecite{Voit2} focused purely on the effect of the el-ph interaction on $K_c^{\rm
eff}$.  Above, we have correctly taken into account that the el-ph interaction affects both
$\Delta_s$ and $K_c^{\rm eff}$, both of which in turn affect $\chi_{\rm SS}$.

\subsection{Near half-filling}

\begin{figure}
\includegraphics[width=0.39\textwidth]{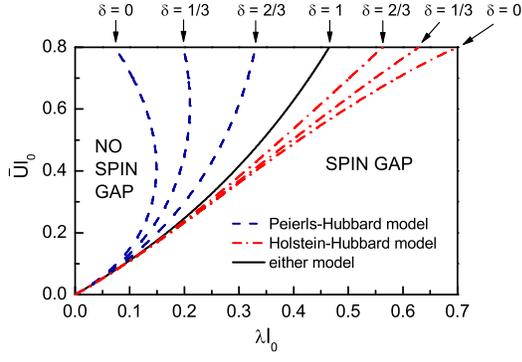}
\caption{\label{bindloss_fig_6} Dependence of the transition line between spin-gapped and
non-spin-gapped phases
on the doping parameter $\delta$ and on the electron-phonon model, for $V = 0$.
For $\mu \le \omega_0$ (which includes the half-filled case $\mu = 0$),
the transition line is denoted by $\delta = 0$.  The incommensurate limit ($\mu \sim E_F$) is
labeled by $\delta = 1$. 
$\delta = 1/3$ and 2/3 are intermediate dopings, with $\mu/\omega_0 = (E_F/\omega_0)^\delta$.
Here, $\lambda$ stands for $\lambda_{\rm Pei}$ in the Pei-Hub model (dashed lines)
and for $\lambda_{\rm Hol}$ in the Hol-Hub model (dashed-dotted lines).
For $\delta = 1$ (solid line), the transition line is the same for either model.
}
\end{figure}

\begin{figure}
\includegraphics[width=0.44\textwidth]{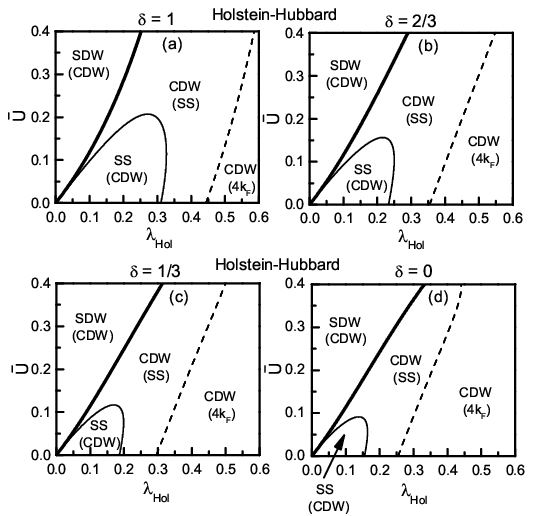}
\caption{\label{Hol_doping} Doping dependence of the Hol-Hub phase diagram
for $V = 0$ and $E_F/\omega_0 = 5$.  The value of $\delta$ is given above each plot.
}
\end{figure}

\begin{figure}
\includegraphics[width=0.44\textwidth]{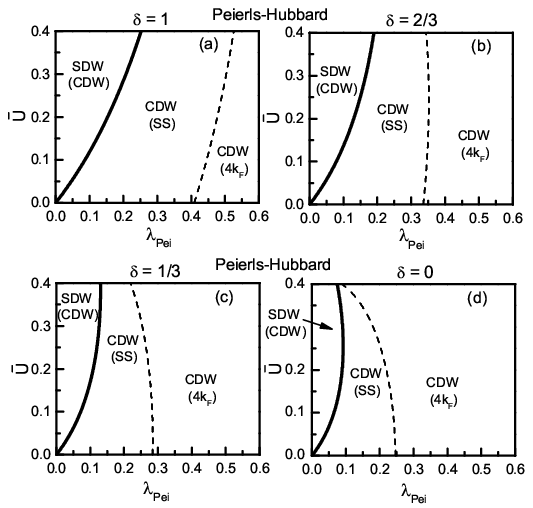}
\caption{\label{Pei_doping} Doping dependence of the Pei-Hub phase diagram
for $V = 0$ and $E_F/\omega_0 = 5$.  The value of $\delta$ is given above each plot.
}
\end{figure}

\begin{figure}
\includegraphics[width=0.43\textwidth]{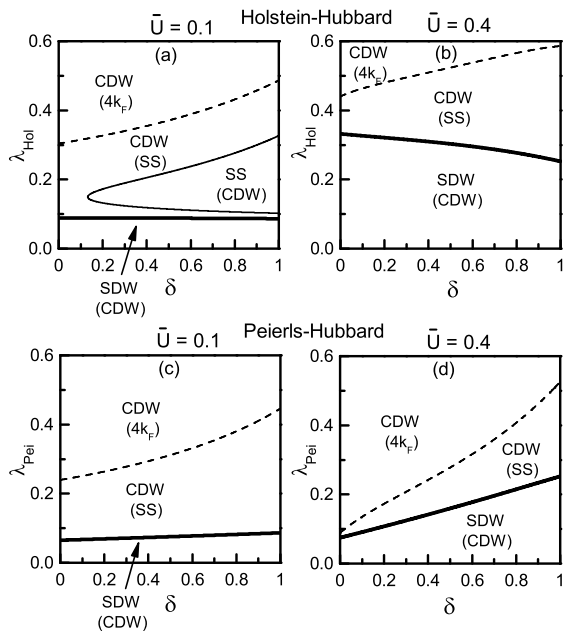}
\caption{\label{fig_1} 
Phase diagrams for $E_F/\omega_0 = 5$ in the $\lambda$-$\delta$ plane.
Plots (a) and (b) are for the Hol-Hub model; (c) and (d) are for the Pei-Hub model.  
Plots (a) and (c) are for ${\bar U} = 0.1$; (b) and (d) are for ${\bar U} = 0.4$.  For all plots,
$V = 0$.
}
\end{figure}

If we reduce the doping level of an incommensurate system
(i.e., move closer to half-filling), the LL-LEL phase boundary is influenced in opposite
ways for the Hol-Hub compared to the
Pei-Hub model.  This is because the on-site Hubbard repulsion 
is in direct competition with the attractive on-site
el-ph interaction of the Holstein model, while it
cooperates with the attractive bond centered interaction of the Pei model.
Therefore, 
the spin gap is enhanced by proximity to
half-filling for the Pei-Hub model,
and reduced for the Hol-Hub model.

We illustrate this in Fig. \ref{bindloss_fig_6}, which presents
a phase diagram in the ${\bar U}l_0$-$\lambda l_0$ plane, for $V = 0$.
This diagram shows the dependence of the LL-LEL phase boundary on the doping
parameter
\be
\delta = \frac{\ln(\mu/\omega_0)}{\ln(E_F/\omega_0)}  .
\label{delta}
\ee
Assuming the charge gap $\Delta_c < \mu$, the actual doping concentration $x$ relative
to half-filling is related to $\delta$ by
\be
x = \frac{2}{\pi v_c} \mu = \frac{2 \omega_0}{\pi v_c}\left(\frac{E_F}{\omega_0}\right)^\delta ,
\ee
where $v_c \approx v_F$ is the charge velocity [see Eq. (\ref{vc})].
The incommensurate limit $\delta = 1$ is
shown previously in
Fig. \ref{bindloss_fig_3}(a) and Fig. \ref{bindloss_fig_3}(b) as a thick solid line.
As we move closer to half-filling by lowering $\delta$, the LL-LEL transition line for
the Pei-Hub model (dashed line) moves toward lower values of $\lambda$, while
the transition line for the Hol-Hub model (dash-dotted line) moves toward higher values.
In the weak-coupling limit assumed here, for $\mu \le \omega_0$, the LL-LEL transition line is
independent of $\mu$.
Therefore, this phase boundary is the same for half-filling ($\mu = x = 0$) as for $\delta = 0$.

The doping dependencies of the other phase boundaries are shown in Figs. \ref{Hol_doping} and
\ref{Pei_doping}, for the range $\omega_0 < \mu < E_F$.  Figure \ref{Hol_doping} illustrates that
for the Hol-Hub model, proximity to half-filling strongly suppresses the phase with dominant SS.
For both models, moving toward half-filling increases the stability of the phase with
subdominant $4 k_F$ charge density wave, at the expense of the phase with subdominant SS,
especially for the Pei-Hub model (Fig. \ref{Pei_doping}).  Note that, especially in Fig.
\ref{Pei_doping}, SS is divergent for a range of parameters such that $\bar U \gg \lambda_{\rm
Pei}$,
despite the low value of $E_F/\omega_0 = 5$.  Figure \ref{fig_1} shows
a different slice of the phase diagram by showing plots in the $\lambda$-$\delta$ plane.

\subsection{Half-filling}

\begin{figure}
\includegraphics[width=0.45\textwidth]{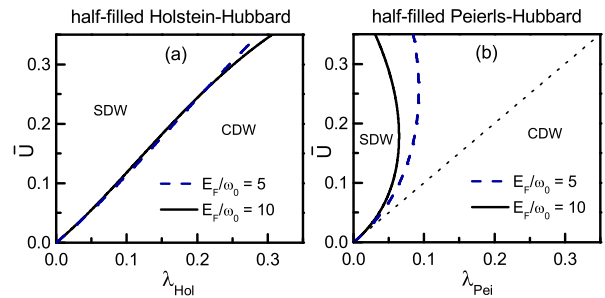}
\caption{\label{bindloss_fig_7} Phase diagram of the (a) half-filled Hol-Hub model
and (b) half-filled Pei-Hub model, both with $V = 0$.
The dashed lines are for $E_F/\omega_0 = 5$ and the solid lines for $E_F/\omega_0 = 10$.
For both diagrams, a spin gap is present for the CDW phase, while a charge gap is present 
everywhere in the phase diagram.  In (b) the dotted line, defined by $\lambda_{\rm Pei} = {\bar
U}$,
is the inaccurate result for the phase boundary from mean-field theory.
}
\end{figure}

At half-filling, for repulsive el-el interactions, a charge gap is present
for the entire phase diagram.  Divergent SS is eliminated at half-filling.
The region without a spin gap is a SDW, while the spin-gapped region is an ordered CDW.
We show the half-filled phase diagram, for both models, in Fig. \ref{bindloss_fig_7},
for several values of $E_F/\omega_0$ and $V = 0$.
The difference between the two models is substantial, with
the Pei-Hub model favoring CDW more than the Hol-Hub model.  In Fig. \ref{bindloss_fig_7}(b),
we also draw a dashed line defined by $\lambda_{\rm Pei} = {\bar U}$, which is the transition
line predicted by mean-field theory.  Such a treatment is known\cite{salkola}
to be quite inaccurate for the Pei-Hub model,
as demonstrated in the figure, since it does not take into account the dramatic renormalization
of the 
backscattering el-ph interaction to stronger couplings.  Note that the Pei-Hub phase
diagram is very sensitive to $E_F/\omega_0$, with high values favoring a spin-gapped CDW,
while the Hol-Hub phase diagram is only weakly dependent on $E_F/\omega_0$.
It is interesting that in the Pei-Hub model [Fig. \ref{bindloss_fig_7}(b)],
there is a maximum value of 
the critical el-ph coupling of about $0.149/l_0$,
which occurs at ${\bar U} \approx 0.411/l_0$.  (In other words, for $\lambda_{\rm Pei} >
0.149/l_0$,
the system is an ordered CDW for any ${\bar U}$.)

\section{Doping dependence of the superconducting susceptibility and isotope effects}
\label{isotopesection}

In this section, we study the strong doping dependencies of the spin gap, superconducting
susceptibility, CDW susceptibility, and isotope effects.

Examining the phase diagram in Fig. \ref{fig_1}(d), we can deduce an interesting nonmonotonic
dependence of the SS susceptibility on $\delta$.
For moderate values of $\lambda_{\rm Pei}$,
for example, near $\lambda_{\rm Pei} \approx 0.2$, $\chi_{\rm SS}$
is not divergent near $\delta = 0$, where only the $2 k_F$ and $4 k_F$ CDW susceptibilities
diverge, nor is it divergent near $\delta = 1$, where
the system is in the gapless LL phase.
However, $\chi_{\rm SS}$ is
divergent for a certain range of moderate $\delta$.
Therefore, in these cases, at fixed $T \ll \Delta_s$, $\chi_{\rm SS}$ must exhibit a maximum as a
function of $\delta$ at some intermediate value of $\delta$.

\begin{figure}
\includegraphics[width=0.41\textwidth]{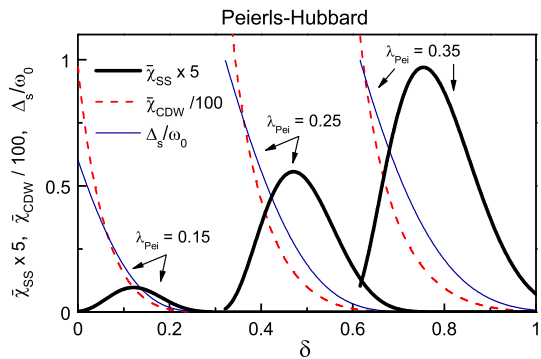}
\caption{\label{fig_2} 
Doping dependence of the
singlet superconducting susceptibility ${\bar \chi}_{\rm SS}$ (thick solid lines) at
$T/\omega_0 = 0.1$,
CDW susceptibility ${\bar \chi}_{\rm CDW}$
(dashed lines) at $T/\omega_0 = 0.1$, and spin gap (thin solid lines), for the Pei-Hub model with
${\bar U} = 0.4$, $V = 0$,
$E_F/\omega_0 = 5$,
and various values of $\lambda_{\rm Pei}$ (labeled in plot).
}
\end{figure}

\begin{figure}
\includegraphics[width=0.475\textwidth]{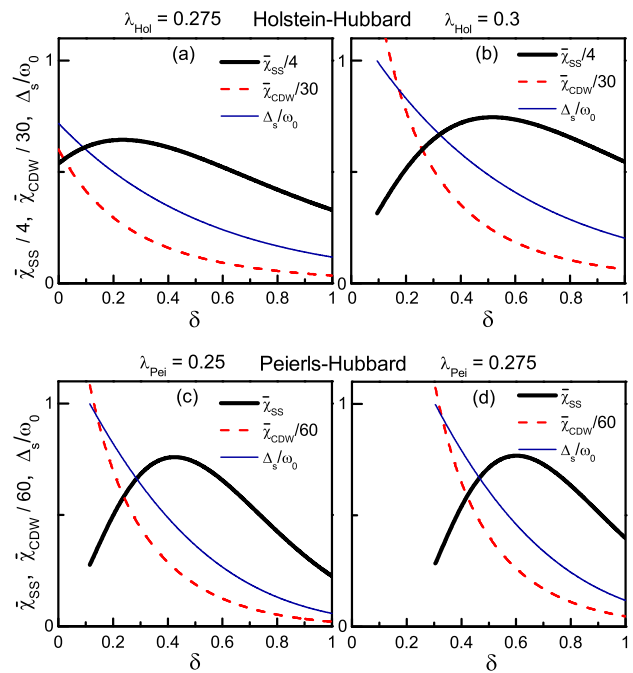}
\caption{\label{fig_3} 
Doping dependence of ${\bar \chi}_{\rm SS}$ (thick solid lines) at $T/\omega_0 = 0.1$,
${\bar \chi}_{\rm CDW}$ (dashed lines) at $T/\omega_0 = 0.1$,
and $\Delta_s/\omega_0$ (thin solid lines), for the Hol-Hub model [(a) and (b)],
and the Pei-Hub model [(c) and (d)].  For all plots, ${\bar U} = 0.1$, $V = 0$, and $E_F/\omega_0
= 5$.
The values of $\lambda_{\rm Hol}$ and $\lambda_{\rm Pei}$
are labeled above each plot.
}
\end{figure}

This maximum in $\chi_{\rm SS}$, which occurs in both models, is shown explicitly in Figs.
\ref{fig_2} and \ref{fig_3},
where we plot ${\bar \chi}_{\rm SS}$ (thick solid line) versus $\delta$
at $T/\omega_0 = 0.1$, for representative parameters.  The cause of the nonmonotonic doping
dependence is the different doping dependencies of $\Delta_s$ and $K_c^{\rm eff}$.  $\Delta_s$
decreases with increasing doping, which acts to reduce $\chi_{\rm SS}$, while $K_c^{\rm eff}$
increases with increasing doping, which acts to increase $\chi_{\rm SS}$.  These two effects
``compete'' with each other and can cause a maximum at some ``optimal'' value of the doping
that depends on the interaction strengths.    
The dimensionless $2 k_F$ CDW susceptibility
\be
\label{barchiCDW}
{\bar \chi}_{\rm CDW} \equiv \pi v_F \chi_{\rm CDW} = (\Delta_s/E_F) (T/E_F)^{K_c^{\rm eff} - 2}
\ee
(dashed lines in Figs. \ref{fig_2} and \ref{fig_3}) does not exhibit such a maximum,
but instead decreases monotonically with increasing doping.
In Figs. \ref{fig_2} and \ref{fig_3}, we also
plot $\Delta_s/\omega_0$ (thin solid lines), which shows that
at low dopings,
$\chi_{\rm SS}$ increases with increasing doping,
despite the fact that the superconducting pairing strength $\Delta_s$ decreases.

We now consider the doping dependence of $T_c$ and the isotope effect on $T_c$ for a quasi-1D
system that consists of an array of weakly coupled quasi-1D chains.
We assume that the chains are spatially or dynamically fluctuating so that CDW order is dephased.
The interchain Josephson coupling $J$ is treated
on a mean-field level,\cite{arrigoni} so that $T_c$ is determined by the temperature at which
\be
\label{determineTc}
2 J \chi_{\rm SS} = 1 \;\;\;\; ({\rm at} \; T = T_c),
\ee
where the numerical prefactor 2 is determined by the number of nearest-neighbor chains.  Treating
the interchain coupling with perturbative RG gives an equivalent result.  
Assuming $J$ is doping independent, $T_c$ then exhibits a maximum at the same $\delta$
where $\chi_{\rm SS}$ has a maximum. 

\begin{figure}
\includegraphics[width=0.495\textwidth]{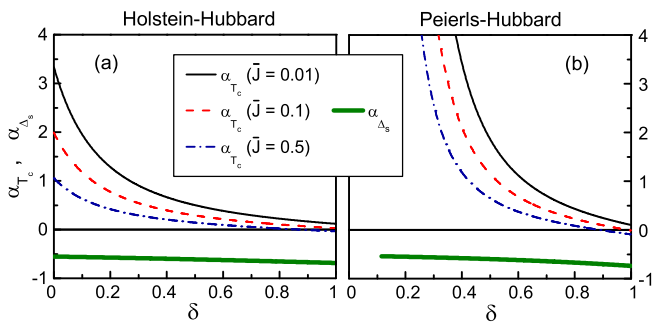}
\caption{\label{fig_4} 
Dependence of the isotope effect exponents $\alpha_{T_c}$ and
$\alpha_{\Delta_s}$ on the doping parameter $\delta$ and 
interchain coupling strength ${\bar J}$
for the Hol-Hub model with $\lambda_{\rm Hol} = 0.275$ (a) and
Pei-Hub model with $\lambda_{\rm Pei} = 0.25$ (b).
For both plots, ${\bar U} = 0.1$, $V = 0$, and $E_F/\omega_0 = 5$.
$\alpha_{\Delta_s}$ is independent of ${\bar J}$.
}
\end{figure}

The isotope effect exponent $\alpha_{T_c}$ is plotted versus $\delta$ in Fig. \ref{fig_4}.
It is shown for various values of ${\bar J} \equiv J/\pi v_F$,
at fixed $\lambda$ and ${\bar U}$.  Unlike in BCS theory, $\alpha_{T_c}$
is not universal but depends on the interaction strengths and band-filling.
However, qualitatively, it appears that the doping-dependent behavior in which  
$\alpha_{T_c}$ is large near half-filling
but decreases rapidly with increasing doping is generic (independent of interaction
strengths and el-ph model). 
Note that if the parameters are tuned just right,
$\alpha_{T_c}$ vanishes.  A small $\alpha_{T_c}$ can even
occur at the doping for which $T_c$ is maximum.
Therefore, one should be careful not to assume that phonons are unimportant in unconventional
superconductors for which $\alpha_{T_c} \ll 1$,
such as in the cuprates at optimal doping.
Figure \ref{fig_4}
also shows $\alpha_{\Delta_s}$, which is
weakly doping-dependent and negative.  

\section{Comparison with other work}
Below, we compare our results for the phase diagrams to some phase diagrams which have been
previously computed.

\label{compare}

\subsection{Half-filled extended Peierls-Hubbard model}

\begin{figure}
\includegraphics[width=0.33\textwidth]{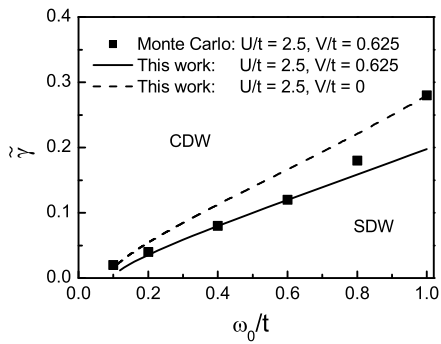}
\caption{\label{bindloss_fig_8} 
Phase diagram of the half-filled Pei-Hub model,
comparing of our result for the phase boundary (solid line)
to the quantum Monte Carlo result of Ref. \onlinecite{Sengupta} (squares),
for $U/t = 2.5$ and $V/t = 0.625$.
The dashed line is our result for the same $U/t$ but with $V = 0$.
Here, $\tilde \gamma  = \sqrt{\pi (\omega_0/t) \lambda_{\rm Pei}}/4$ is the el-ph coupling
constant in Eq. (\ref{Pei}).
}
\end{figure}

We first compare our results for
the half-filled extended Pei-Hub model to
recent Monte Carlo work by Sengupta, Sandvik, and Campbell\cite{Sengupta} on the same model.
Following Ref. \onlinecite{Sengupta}, we show a phase diagram in the $\tilde \gamma$-$\omega_0/t$
plane in Fig. \ref{bindloss_fig_8}.  This figure compares our result for the critical line 
for $U/t = 2.5$ and $V/t = 0.625$ (solid line)
to the result in Ref. \onlinecite{Sengupta}.
(In order to plot our
result in these units, we took the high-energy cutoff $E_F$ in the RG theory to be $t$.)

The quantitative disagreement near $\omega_0/t \sim 1$
can probably be attributed to the fact that
the assumption in the RG theory of a linear electronic dispersion
becomes problematic when $\omega_0 \sim E_F$.
At lower $\omega_0/t$, the agreement is excellent, especially considering
the moderately strong value ${\bar U} = U/(2 \pi t) \approx 0.4$.
This gives one reason to believe that the multistep RG method is,
at the very least, qualitatively accurate for physically interesting values
of ${\bar U} \sim 1$.
 
In Fig. \ref{bindloss_fig_8}, we have also plotted our result for $V = 0$ (dashed line),
which can be obtained from the 
simple analytic expression in Eq. (\ref{13}).
The solid line in this figure is the only phase boundary in the present paper
that required numerical integration of the RG flow equations
[Eqs. (\ref{5}) and (\ref{6})].
The phase boundaries in all other plots are given by
analytic expressions derived in Section \ref{technique}.
 
\subsection{Holstein model}

\begin{figure}
\includegraphics[width=0.25\textwidth]{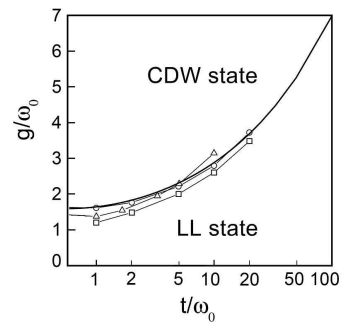}
\caption{\label{bindloss_fig_9} Phase diagram of the spinless half-filled
Holstein model from various authors, in the absence of el-el interactions.  The solid line is an
analytical result from Ref. \onlinecite{Wang}.
The line with circles and the line with squares denote the results of density matrix
renormalization group (Ref. \onlinecite{Bursill}) and two-step RG (Ref. \onlinecite{Caron}),
respectively.  The line with triangles
is the result from an exact-diagonalization method (Ref. \onlinecite{Weisse}).  (After Ref.
\onlinecite{Wang}.)}
\end{figure}

The phase diagram for perhaps the most interesting model
studied in the present paper,
the spinful incommensurate extended Holstein-Hubbard model,
has not extensively studied in prior works.
A much simpler related model, which has been thoroughly explored,
is the spinless half-filled
Holstein model (without el-el interactions).  We show the phase diagram for this model
in Fig. \ref{bindloss_fig_9}, computed by various authors with a wide
range of methods.  Note that the result from the 
two-step RG technique (line with squares)\cite{Caron}
is in good agreement with exact numerical methods.

\begin{figure}
\includegraphics[width=0.255\textwidth]{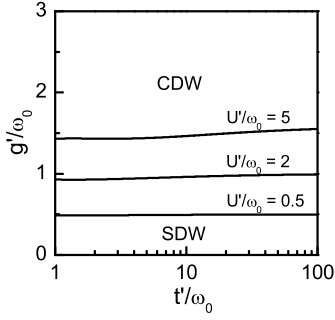}
\caption{\label{bindloss_fig_10} Phase diagram in the $g^\prime/\omega_0$-$t^\prime/\omega_0$
plane of the spinful half-filled Holstein-Hubbard model for $V = 0$.  The phase boundary
is shown for various values of $U^\prime/\omega_0$, as labeled.
The system is charge-gapped
everywhere in the phase diagram, and spin-gapped in the CDW phase.
For $U^\prime/\omega_0 = 0$ (not shown), the ground state is
a CDW for infinitesimal $g^\prime/\omega_0$.
Here, $g^\prime \approx g$,
$t^\prime \approx t$, and $U^\prime \approx U$ (see the text for their precise definitions).
}
\end{figure}

In order to see how Fig. \ref{bindloss_fig_9} changes
when spin is included, we have computed a phase diagram
in similar units for
the spinful half-filled Holstein model in Fig. \ref{bindloss_fig_10},
including a Hubbard interaction.
In the absence of el-el interactions, at weak el-ph coupling,
the ground state of this model is always a CDW, for
any finite $\omega_0$.
This also holds in the strong
coupling limit ($\lambda_{\rm Hol} \gg 1$).\cite{Fradkin2}
In contrast, for the {\it spinless} half-filled Holstein model
without el-el interactions, the transition to a CDW occurs at a nonzero value of $g$,
as shown in Fig. \ref{bindloss_fig_9} for weak coupling and proven in Ref. \onlinecite{Fradkin2}
for strong coupling.

To study how Fig. \ref{bindloss_fig_10} changes
when the system is doped into the incommensurate limit,
we have presented a diagram in the same units for
the spinful incommensurate extended Holstein-Hubbard model in Fig. \ref{bindloss_fig_5}.
In Figs. \ref{bindloss_fig_5} and \ref{bindloss_fig_10},
for technical reasons, we have defined $g^\prime = g \sqrt{E_F/v_F}$,
$t^\prime = t(E_F/2t)$, and $U^\prime = U(E_F/v_F)$.

\section{Methods and derivations}
\label{technique}

Here, we provide more discussion of the technique and
study the RG flows of the coupling parameters.  We also derive
the explicit expressions that were used to plot the phase 
boundaries in the figures.

\subsection{Field theory for the 1DEG coupled to phonons}

To focus on the low-energy, long-wavelength physics,
we work with continuum versions of Eqs.
(\ref{Pei}) and (\ref{Hol}).  The purely electronic part of our Hamiltonian is the standard
continuum model of
the interacting 1DEG, in which the spectrum is linearized around the left and right Fermi points.  
The
destruction field for fermions of spin $\sigma$ is written as a sum of slowly varying right and
left moving fields: $\Psi_{\sigma
} = e^{ik_Fx}\psi_{1,\sigma} + e^{-ik_Fx}\psi_{-1,\sigma}$.  The
Hamiltonian density for the 1DEG, in the absence of el-ph interactions, is written as
${\cal H} = {\cal H}_0 + {\cal H}_{\rm el-el}$, where the kinetic energy density is
\be
{\cal H}_0 = -i v_F \sum_{\eta,\sigma = \pm1} \eta \, \psi_{\eta,\sigma}^\dagger  \partial_x
\psi_{\eta,\sigma},   
\label{H_0}
\ee
and the important short range el-el interaction terms are
\ba
\nonumber {\cal H}_{\rm el-el} &=& g_1 \sum_{\sigma,\sigma^\prime = \pm1} \psi_{1,\sigma}^\dagger
\psi_{-1,\sigma^\prime}^\dagger \psi_{1,\sigma^\prime} \psi_{-1,\sigma} \\
\nonumber &+& g_2 \sum_{\sigma,\sigma^\prime = \pm1} \psi_{1,\sigma}^\dagger
\psi_{-1,\sigma^\prime}^\dagger \psi_{-1,\sigma^\prime} \psi_{1,\sigma} \\
\nonumber &+& g_3 \, [e^{i(4k_F - G)x} \, \psi_{-1,1}^\dagger \psi_{-1,-1}^\dagger \psi_{1,-1}
\psi_{1,1} + {\rm H.c.}]\\
&+& g_4 \sum_{\eta,\sigma = \pm1} \psi_{\eta,\sigma}^\dagger \psi_{\eta,-\sigma}^\dagger
\psi_{\eta,-\sigma} \psi_{\eta,\sigma} .
\label{H_el-el}
\ea
We have assumed the system is spin-rotation invariant.
The $g_2$  and $g_4$ terms describe
forward scattering,
the former containing scattering on both left and right moving branches, and the latter
containing scattering
on only one branch.  The $g_1$ term contains backscattering from one branch to the other. The
$g_3$  term
contains Umklapp processes and is only important when $4 k_F$ equals a reciprocal lattice vector
$G$; i.e., at half-filling ($4 k_F = 2 \pi$).
For the extended Hubbard model in the continuum limit,
the bare (unrenormalized) values of the $g_i$'s are given by
\ba
\label{extendedHubbard}g_1^0 &=& g_3^0 \:=\: U - 2V^\prime , \\
g_2^0 &=& U + 2V^\prime  , \\
g_4^0 &=& U/2 + 2V^\prime 
\ea
where the superscript $0$ indicates bare couplings and again   
$V^\prime = -V \cos(2 k_F)$, which equals $V$ at half-filling.

We incorporate el-ph interactions by defining retarded interactions
$g_{1,{\rm ph}}$, $g_{2,{\rm ph}}$, $g_{3,{\rm ph}}$, and $g_{4,{\rm ph}}$ which play the
same role as the $g_i$'s except that the energy transfer is restricted to be less than a
cutoff $\omega_c$, which is approximately the phonon frequency $\omega_0$ ($\omega_c$ will
be defined more precisely below).
This corresponds to approximating the phonon propagator as a step function of frequency, which is
a
good approximation for the momentum-independent phonon dispersions we consider in this paper.
In the Pei-Hub model, the bare el-ph couplings are given by
\ba
\label{Pei_gs}
g_{1,{\rm ph}}^0 &=& -g_{3,{\rm ph}}^0 \:=\: - \pi v_F \lambda_{\rm Pei} , \\
g_{2,{\rm ph}}^0 &=&  g_{4,{\rm ph}}^0 \:=\: 0  .
\ea
For the Hol-Hub model they are
\be
\label{Holstein_gs}
g_{1,{\rm ph}}^0 =  g_{2,{\rm ph}}^0 = g_{3,{\rm ph}}^0 =  g_{4,{\rm ph}}^0 = - \pi v_F
\lambda_{\rm Hol} ,  
\ee
where $\lambda_{\rm Pei}$ and $\lambda_{\rm Hol}$ are the positive, dimensionless coupling
constants
defined in Eq. (\ref{couplingconstants}).  Note that the couplings $g_{1,{\rm ph}}^0$,
$g_{2,{\rm ph}}^0$, and $g_{4,{\rm ph}}^0$ are negative, indicating the
attractive interaction induced by phonons.
In the absence of el-el interactions,
the sign of $g_{3,{\rm ph}}^0$ is arbitrary; likewise, the sign
of $g_3^0$ is arbitrary in the absence of el-ph interactions.
However, for the extended Hubbard model coupled to phonons, 
once the sign convention $g_3^0 = U - 2 V^\prime$ is chosen,
it is required that
$g_{3,{\rm ph}}^0 > 0$ for the Pei-Hub model and
$g_{3,{\rm ph}}^0 < 0$ for the Hol-Hub model.

\subsection{Phase diagram of the 1DEG without phonons}

Before deriving the phase diagram including phonons,
we briefly review the known quantum phase
diagram of the 1DEG without el-ph coupling,\cite{giamarchi,giamarchiPaper,KivelsonBook} by
identifying the conditions for various types of order to have divergent susceptibilities 
in the low-temperature limit.

The sign of $g_1^0$ determines the existence or nonexistence of a spin gap:
a 1DEG without el-ph coupling contains a gap to spin excitations for $g_1^0 < 0$,
and no such gap for $g_1^0 \ge 0$, regardless of band filling.
A charge gap is only possible at commensurate fillings.

\begin{table}
\caption{\label{table1}Conditions for various
ordering fluctuations to dominate for an incommensurate
1DEG.  See the text for the subdominant fluctuations.
}
\begin{ruledtabular}
\begin{tabular}{lcr}
Type of order&Dominates for\footnote{
For the Hubbard model, $K_c < 1$ corresponds to repulsive
interactions, $K_c > 1$ to attractive interactions. The effect of a forward scattering el-ph
interaction is to raise $K_c$, while a backscattering el-ph interaction lowers $K_c$ and, if
strong enough compared to the el-el repulsion, causes a gap in the spin sector ($\Delta_s >
0$).}\\
\hline
$4k_F$ charge density wave ($4 k_F$) & $\Delta_s = 0$, $K_c < 1/3$\\
$2k_F$ spin density wave & $\Delta_s = 0$, $1/3 < K_c < 1$\\
$2k_F$ charge density wave & $\Delta_s > 0$, $K_c < 1$\\
Triplet superconductivity & $\Delta_s = 0$, $K_c > 1$\\
Singlet superconductivity & $\Delta_s > 0$, $K_c > 1$\\
\end{tabular}
\end{ruledtabular}
\end{table}

\begin{table}
\caption{\label{table2}Conditions for SDW and CDW
order in a half-filled, charge-gapped 1DEG.}
\begin{ruledtabular}
\begin{tabular}{lcr}
Type of order&Present for\\
\hline
$2k_F$ spin density Wave & $\Delta_s = 0$\\
$2k_F$ charge Density Wave & $\Delta_s > 0$\\
\end{tabular}
\end{ruledtabular}
\end{table}

An incommensurate 1DEG without a spin gap (Luttinger liquid), has a divergent $2 k_F$
SDW susceptibility when the Luttinger charge exponent 
\be
K_c = \sqrt{\frac{2 \pi v_F + 2 g_4^0 + g_1^0 - 2 g_2^0}{2 \pi v_F + 2 g_4^0 - g_1^0 + 2 g_2^0}}
\ee
is less than 1, along with a logarithmically more weakly divergent
$2 k_F$ CDW correlation.
If $K_c < 1/2$, the $4 k_F$ CDW correlation is also divergent; it is less divergent than
SDW and $2 k_F$ CDW for $1/3 < K_c < 1/2$, but becomes the dominant order for $K_c < 1/3$.
For $K_c > 1$, a LL has a divergent triplet superconducting (TS) correlation and a
logarithmically weaker divergent singlet superconductivity.

An incommensurate spin-gapped 1DEG (Luther-Emery liquid), is dominated by either
SS or $2 k_F$ CDW correlations, depending on the value of $K_c$.
A LEL with $K_c < 1$ is dominated by $2 k_F$ CDW.
If $K_c < 1/2$, $4 k_F$ CDW is subdominant, while if $1/2 < K_c < 1$, 
SS fluctuations replace $4 k_F$ CDW as the subdominant order.  For $1 < K_c < 2$, SS is the most
divergent channel, and CDW is subdominant.  For $K_c > 2$, the only divergent correlations is
SS.
We summarize the phase diagram of the incommensurate 1DEG in Table \ref{table1}.

At half-filling, a 1DEG is charge-gapped for $|g_3^0| > g_1^0 - 2g_2^0$, which for
repulsive interactions is always satisfied.
For $g_1^0 < 0$, the ground state of a charge-gapped system
is an ordered, spin-gapped CDW (the Peierls instability),
otherwise, it is a SDW with no spin gap.

\subsection{The multistep RG technique}

To determine the phase diagrams,
we apply the known one-loop RG flow equations\cite{Steve}
\ba
\label{5}&&\frac{d g_1}{d l} = -g_1^2 , \; \; \: \frac{d g_c}{d l} = -g_3^2 , \; \; \: \frac{d
g_3}{d l} = -g_3 g_c  , \;\;\;\;\;\\
\label{6}&&\frac{d g_{\pm,{\rm ph}}}{d l} = -g_{\pm,{\rm ph}}\left( \frac{3}{2}g_1 \pm g_3 +
\frac{1}{2}g_c + g_{\pm,{\rm ph}}  \right) , \\  
\label{6b}&&\frac{d g_{2,{\rm ph}}}{d l} = \frac{d g_{4,{\rm ph}}}{d l} = \frac{d g_4}{d l} = 0 ,
\\
\label{7}&&\frac{d \omega_c}{d l} = \omega_c (\pi v_F + g_{+,{\rm ph}}) , \; \; \: \frac{d \mu}{d
l}  = \pi v_F \mu ,
\ea
where we defined
\ba
&& \; g_c = g_1 - 2g_2 , \\ 
&& \; g_{\pm,{\rm ph}} = g_{1,{\rm ph}} \pm g_{3,{\rm ph}} ,\\
&& \; l = (\pi v_F)^{-1}\ln(E_F/\omega) , 
\ea
and $\omega$ is the running cutoff.
The above expressions apply for $E_F \gg \omega \gg \omega_0, \mu$.  If, instead, 
$\mu \gg \omega \gg \omega_0$, the same equations apply, but
with $g_3 = g_{3,{\rm ph}} = 0$.
>From Eq. (\ref{5}), we see that a repulsive $g_1$ renormalizes
in the same way as
the Coulomb pseudopotential in a Fermi liquid--it is scaled to smaller values as one integrates
out high-energy degrees of freedom.  
However, in 1D the backscattering and Umklapp el-ph
interactions $g_{1,{\rm ph}}$ and $g_{3,{\rm ph}}$ are strongly renormalized.
Furthermore, there are cross terms
$g_{\pm,{\rm ph}} g_i$  in Eq. (\ref{6}), which means that the RG flows of
$g_{1,{\rm ph}}$ and $g_{3,{\rm ph}}$ are strongly
influenced by direct el-el interactions.

The two-step RG procedure is as follows:
Assuming $\Delta_s < \omega_0$ and the system
is at half-filling ($\mu = 0$), we first integrate out fermionic degrees of freedom between the
high-energy scale
$E_F$ and the phonon energy $\omega_0$ using Eqs. (\ref{5}) and (\ref{6}).   
Once $\omega_0$ is reached, total effective interactions $g_i^{\rm tot}(\omega_0)$
at this energy scale are determined by adding the effective el-el coupling to the effective el-ph
coupling
\be
g_i^{\rm tot}(\omega_0) = g_i(\omega_0) + g_{i,{\rm ph}}(\omega_0)  .
\label{8}
\ee
Below this energy scale, there is no difference between retarded and instantaneous interactions,
so $g_i^{\rm tot}$ renormalizes as a non-retarded interaction using Eq. (\ref{5}) with $g_i^{\rm
tot}(\omega_0)$
as the initial value.
If the system is far enough
away from half-filling such that $\mu \sim E_F$, the method is identical except that one sets
$g_3^0 = g_{3,{\rm ph}}^0 = 0$ at the start.
For the above cases, it is clear that
the renormalized couplings $g_i^{\rm tot}(\omega_0)$
and the renormalized cutoff $\omega_0$
play the same
role in the 1D electron-phonon system as $g_i^0$ and $E_F$, respectively, do in the pure 1DEG.  
Therefore, to determine the phase diagram of the 1DEG coupled to phonons, we can use
the known phase diagram of the pure 1DEG, and simply replace the $g_i^0$'s there with
$g_i^{\rm tot}(\omega_0)$'s.
For more general fillings ($0 < \mu < E_F$), a
three-step RG method is necessary, which we will elaborate at the end of this section.

Note that
$\omega_0$ is the physical phonon frequency, which is related to $\omega_c$ by the expression
$\omega_0  = \omega_c(\omega_0)$, where $\omega_c(\omega_0)$ is the renormalized value of
$\omega_c$ at the energy scale $\omega_0$, determined by Eq. (\ref{7}).
Likewise,
we define $\mu$ as the physical value of the chemical potential relative to its value at half-filling;
the bare chemical potential $\mu^0$ is chosen
such that it flows to the value $\mu$
after integrating out degrees of freedom between $E_F$ and $\mu$. 

\subsection{Incommensurate filling}
We first derive the scaling of the coupling constants
and compute the phase diagrams for the case when the
system is doped far into the incommensurate limit
($\mu \sim E_F$).

\subsubsection{Scaling of the coupling constants}

Using Eqs. (\ref{5}) and (\ref{6}) with $g_3 = g_{3,{\rm ph}} = 0$,
we integrate out degrees of freedom between $E_F$  and $\omega$
with $E_F > \omega \ge \omega_0$,
to obtain the follow effective couplings for incommensurate systems:
\ba
g_c(\omega) &=& g_c^0 , \\
\label{g1omega} g_1(\omega) &=& \frac{g_1^0}{1 + g_1^0 l}  , \\
\nonumber g_{1,{\rm ph}}(\omega) &=& \left( \frac{g_{1,{\rm ph}}^0}{1 + g_{1,{\rm ph}}^0
Z}\right)\left(\frac{g_1(\omega)}{g_1^0}\right)^{3/2}\left(\frac{E_F}{\omega}\right)^{-g_
c^0 / 2 \pi v_F} , \\ 
\label{g1omegaph}
\ea
where
\be
Z = \int_0^l dx \, \frac{\exp(-g_c^0 x/2 )}{(1 + g_1^0 x)^{3/2}},
\ee
and again $g_i^0 = g_i(E_F)$.

\begin{figure}
\includegraphics[width=0.44\textwidth]{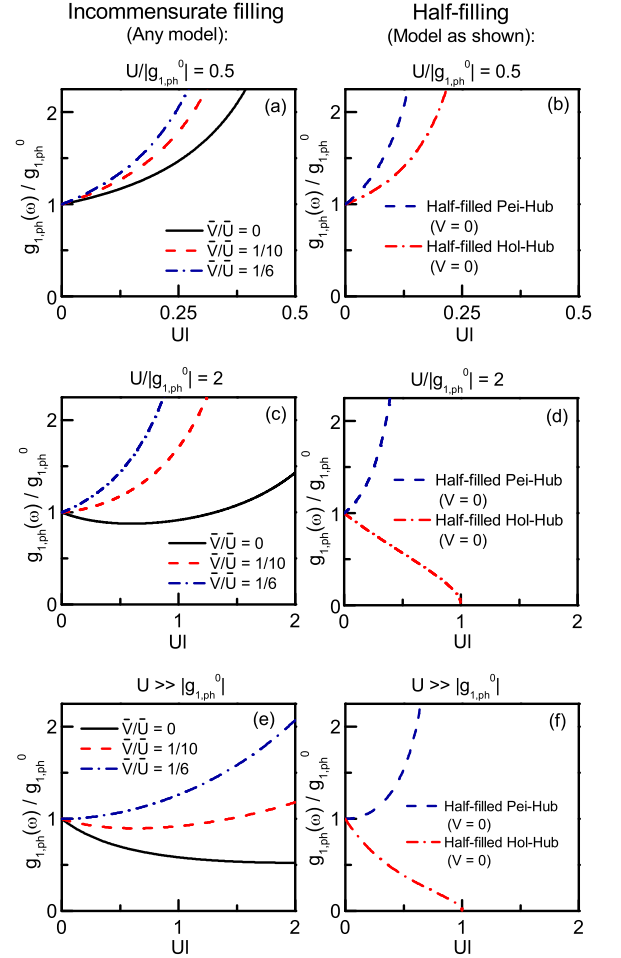}
\caption{\label{bindloss_fig_11} 
Dependence of the effective backscattering el-ph coupling $g_{1,{\rm ph}}(\omega)$ 
on $Ul \equiv (U/\pi v_F) \ln(E_F/\omega)$
for $\omega > \omega_0$.
Plots (a), (c), and (e) are
for systems doped into the incommensurate limit, while
(b), (d) and (f) are for
the half-filled Pei-Hub and Hol-Hub models with $V = 0$. 
For each plot, the ratio of $U$
to the bare el-ph coupling $g_{1,{\rm ph}}^0$
is held fixed at the value indicated above the plot.
For each curve in (a), (c), and (e), the ratio ${\bar V}/{\bar U}$ is held fixed at the value
indicated.
}
\end{figure}

\begin{figure}
\includegraphics[width=0.44\textwidth]{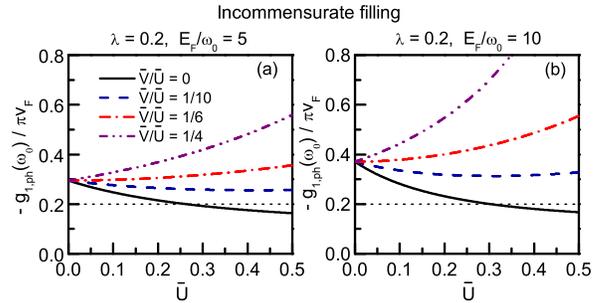}
\caption{\label{versusU} 
Dependence of $g_{1,{\rm ph}}(\omega_0)$ 
on ${\bar U}$ for incommensurate systems
with (a) $E_F/\omega_0 = 5$ and (b) $E_F/\omega_0 = 10$ (b).  The horizontal dotted line
indicates the value of the bare el-ph coupling $\lambda = - g_{1,{\rm ph}}^0/\pi v_F = 0.2$,
where $\lambda$ stands for either $\lambda_{\rm Hol}$ or $\lambda_{\rm Pei}$,
depending on the model.  For each curve, the ratio ${\bar V}/{\bar U}$ is held fixed at the value
indicated in (a).
}
\end{figure}

In the absence of el-el interactions, the 
el-ph backscattering coupling $g_{1,{\rm ph}}$ flows to stronger values
according to $g_{1,{\rm ph}}(\omega)/g_{1,{\rm ph}}^0 = (1 - |g_{1,{\rm ph}}^0|l )^{-1}$.
In Fig. \ref{bindloss_fig_11},
the important influence of extended Hubbard interactions
on $g_{1,{\rm ph}}(\omega)$ in studied.
In Figs. \ref{bindloss_fig_11}(a) and \ref{bindloss_fig_11}(c), we
show the scaling
for several fixed finite values of $U/|g_{1,{\rm ph}}^0|$.
Fig. \ref{bindloss_fig_11}(e) shows the scaling 
in the limit $U \gg |g_{1,{\rm ph}}^0|$, which is given by
\be
g_{1,{\rm ph}}(\omega)/g_{1,{\rm ph}}^0 \approx h(Ul),  \;\;\;\;\;\; (U \gg |g_{1,{\rm ph}}^0|) , 
\ee
where for future convenience we define
\ba
\label{hx}
h(x) &=& \frac{\exp(G_c x/2)}{(1 + G_1x)^{3/2}}  , \\
G_1 &=& 1 - 2({\bar V}/{\bar U})  , \\
G_c &=& 1 + 6({\bar V}/{\bar U})  .
\ea

>From Fig. \ref{bindloss_fig_11}, we see that the flow of $g_{1,{\rm ph}}$ is
very sensitive to the parameter ${\bar V}/{\bar U}$.
If ${\bar V}/{\bar U} < 1/6$ and $U/|g_{1,{\rm ph}}^0|$ is large enough,
then the $g_{\pm,{\rm ph}}g_1$ term in Eq. (\ref{6})
causes $g_{1,{\rm ph}}(\omega)$ to initially flow to {\it weaker} values.
However, if ${\bar V}/{\bar U} \ge 1/6$, then
$g_{1,{\rm ph}}(\omega)/g_{1,{\rm ph}}^0 > 1$ for all $\omega < E_F$,
regardless of $U/|g_{1,{\rm ph}}^0|$.   
The driving force for this increase in $|g_{1,{\rm ph}}(\omega)|$
is the $g_{\pm,{\rm ph}}g_c$ term in Eq. (\ref{6}).

To better understand why increasing ${\bar U}$ can sometimes
enhance superconductivity at small ${\bar U}$ (see Fig. \ref{increaseU}), we
study the dependence of $g_{1,{\rm ph}}(\omega_0)$ on ${\bar U}$ at fixed $E_F/\omega_0$
in Fig. \ref{versusU}.  In this plot, for small ${\bar U}$,  
the effective $g_{1,{\rm ph}}$ is stronger that its bare value ($g_{1,{\rm
ph}}(\omega_0)/g_{1,{\rm ph}}^0 > 1$).  However, for ${\bar V}/{\bar U} < 1/6$, increasing ${\bar U}$
causes $|g_{1,{\rm ph}}(\omega_0)|$ to decrease.
This can cause $K_c^{\rm eff}$ and $\chi_{\rm SS}$ to increase, as in Fig. \ref{increaseU}.
However, if we set ${\bar V}/{\bar U} \ge 1/6$ and hold this ratio fixed while increasing ${\bar
U}$, then $|g_{1,{\rm ph}}(\omega_0)|$ increases, causing $K_c^{\rm eff}$
and $\chi_{\rm SS}$ to be suppressed.

Note that in real materials, 
one typically expects ${\bar V}/{\bar U} > 1/6$,
in which case $|g_{1,{\rm ph}}(\omega_0)| > |g_{1,{\rm ph}}^0|$
or even $|g_{1,{\rm ph}}(\omega_0)| \gg |g_{1,{\rm ph}}^0|$.
The requirement for a spin gap is $|g_{1,{\rm ph}}(\omega_0)| > g_1(\omega_0)$,
which can be achieved, in many cases, with even a small amount of retardation.
(Note that for repulsive el-el interactions, $0 < g_1(\omega_0) < g_1^0$.)
Therefore, it is possible for a slightly retarded el-ph interaction to create 
a divergent superconducting susceptibility
even when the bare interactions are predominantly repulsive
($|g_{1,{\rm ph}}^0| \ll g_1^0$).

\subsubsection{Luttinger liquid to Luther-Emery liquid transition}

The phase boundary between the LL and LEL phases is given by the condition $g_1^{\rm
tot}(\omega_0) = 0$.  This condition determines the following
critical value of the bare el-ph coupling
\ba
\label{spingapeq} \lambda^{\rm Gap}_{\rm Pei} = \lambda^{\rm Gap}_{\rm Hol} &=& {\bar
U}\left[\frac{\exp(G_c {\bar U}l_0/2)}{G_1\sqrt{1 + G_1{\bar U}l_0}} + f({\bar U}l_0)\right]^{-1}
\;\;\;\;\\
\nonumber \\
\nonumber &&{\rm (LL{\rm -}LEL \; transition)},
\ea
where we define
\be
f(y) = \int_0^y dx \, h(x) .
\label{fy}
\ee
For $\lambda_{\rm Pei} > \lambda^{\rm Gap}_{\rm Pei}$ or
$\lambda_{\rm Hol} > \lambda^{\rm Gap}_{\rm Hol}$, the
ground state of the incommensurate 1DEG is a spin-gapped LEL; otherwise,
it is a gapless LL.
The condition $\lambda = \lambda^{\rm Gap}_{\rm Pei} = \lambda^{\rm Gap}_{\rm Hol}$
therefore determines the
transition lines in Figs. \ref{bindloss_fig_1} and \ref{bindloss_fig_2},
as well as the thick solid lines in Figs. \ref{bindloss_fig_3}, \ref{bindloss_fig_4}, and
\ref{bindloss_fig_5}. 
Note that in Fig. \ref{bindloss_fig_2}, we used that
the ``scaled'' critical couplings $\lambda^{\rm Gap}_{\rm Hol} l_0$
and $\lambda^{\rm Gap}_{\rm Pei} l_0$
depend only on the two parameters ${\bar U}l_0$ and ${\bar V}/{\bar U}$.

\subsubsection{Susceptibilities and spin gap in the LEL phase}

In the LEL phase, the potentially strongly divergent part of the low-temperature 
susceptibilities for SS and $2 k_F$ CDW are given by Eqs. (\ref{barchiSS}) and (\ref{barchiCDW}),
respectively, where
$K_c^{\rm eff}$ is the effective Luttinger charge exponent after integrating out states
between $E_F$ and $\omega_0$, given by
\be
\label{15}
K_c^{\rm eff} = \sqrt{\frac{2\pi v_F + 2g_4^{\rm tot} + g_c^{\rm tot}(\omega_0)}{2\pi v_F +
2g_4^{\rm tot} - g_c^{\rm tot}(\omega_0)}},
\ee
where
\ba
\label{16} g_c^{\rm tot}(\omega_0) &=& g_c^0 + g_{1,{\rm ph}}(\omega_0) - 2g_{2,{\rm ph}}^0  , \\
\label{17} g_4^{\rm tot} &=& g_4^0 + g_{4,{\rm ph}}^0  .
\ea
Integrating out states
below the energy $\omega_0$ does not further renormalize
$g_c^{\rm tot}$ (and therefore $K_c$).  Note that
the effective charge and spin velocities
are also renormalized due to phonons, the former of which is
\be
\label{vc}
v_c = (2\pi)^{-1} \sqrt{[2\pi v_F + 2g_4^{\rm tot}]^2 - [g_c^{\rm tot}(\omega_0)]^2} .
\ee

For $-1 < g_1^{\rm tot}(\omega_0) < 0$, $\Delta_s$
is given approximately by the energy scale below $\omega_0$ at which the
RG analysis breaks down because the effective
$g_1^{\rm tot}/\pi v_F$ has grown to $-1$.
Since
\be
\nonumber g_1^{\rm tot}(\omega) = \frac{g_1^{\rm tot}(\omega_0)}{1 + [g_1^{\rm tot}(\omega_0)/\pi
v_F]\ln(\omega_0/\omega)} \;\;\;\;\;{\rm for} \; \omega < \omega_0 ,
\ee
this gives
\be
\label{computedelta}
\Delta_s = \omega_0 e \exp[-\pi v_F/|g_1^{\rm tot}(\omega_0)|]  .
\ee
For $g_1^{\rm tot}(\omega_0) > 0$,
$g_1^{\rm tot}(\omega) \rightarrow 0$ as $\omega \rightarrow 0$; therefore, $\Delta_s = 0$. 

\subsubsection{Competition between SS and CDW in the Hol-Hub model}

The thin solid line in the incommensurate extended Hol-Hub model phase diagrams of
Figs. \ref{bindloss_fig_3}, \ref{bindloss_fig_4}, and \ref{bindloss_fig_5}, 
which we call the ``superconducting transition,'' is determined by $K_c^{\rm eff} = 1$.
This condition is satisfied for $\lambda_{\rm Hol} = \lambda^{{\rm SS}, +}_{\rm Hol}$
and $\lambda_{\rm Hol} = \lambda^{{\rm SS}, -}_{\rm Hol}$ where
\ba
&&\label{18} \lambda^{{\rm SS}, \pm}_{\rm Hol} = {\bar U}\left[ H \pm \sqrt{H^2 - \frac{G_c}{2
f({\bar U}l_0)}} \right]\;\;\;\;\\
\nonumber\\
\nonumber&& \; \;\; \; \; \;  {\rm (superconducting \; transition)}
\ea
and
\be
H = \frac{2 - h({\bar U}l_0)}{4 f({\bar U}l_0)} + \frac{G_c}{4}  .
\ee
Therefore, as long as the square root is not imaginary, for fixed ${\bar U}l_0 > 0$,
there are {\it two} critical values of the bare el-ph coupling determining the boundary between
the phase
with dominant CDW and the phase with dominant SS.
The most divergent correlation is SS provided
$G_c < 2 f({\bar U}l_0)H^2$ and
$\lambda^{{\rm SS}, -}_{\rm Hol} < \lambda_{\rm Hol} < \lambda^{{\rm SS}, +}_{\rm Hol}$.
Note that divergent a TS susceptibility is not present in any phase diagrams,
since for repulsive el-el interactions, $K_c^{\rm eff} < 1$ whenever $g_1^{\rm tot}(\omega_0) >
0$.
  
The phase boundary shown as a dashed line in Figs. \ref{bindloss_fig_3},
\ref{bindloss_fig_4}, and \ref{bindloss_fig_5}
is defined by $K_c^{\rm eff} = 1/2$.
For the extended Hol-Hub model, this condition
occurs at $\lambda_{\rm Hol} = \lambda^{\rm CDW}_{\rm Hol}$
with
\ba
&&\label{21} \lambda^{\rm CDW}_{\rm Hol} = {\bar U}\left[ L + \sqrt{L^2 + \frac{Q}{2 f({\bar
U}l_0)}} \right]\;\;\;\; \\
\nonumber\\
\nonumber&& \;\;\;\; {\rm(} K_c^{\rm eff} = 1/2 \; {\rm transition),}
\ea
where
\ba
L &=& \frac{4 - 5 h({\bar U}l_0)}{8 f({\bar U}l_0)} - \frac{Q}{4} , \\
Q &=& 3/{\bar U} - 9({\bar V}/{\bar U}) - 1  .
\ea
For $\lambda_{\rm Hol} > \lambda^{\rm CDW}_{\rm Hol}$,
a divergent SS susceptibility is not possible.

\subsubsection{Competition between SS and CDW in the Pei-Hub model}

For the extended Pei-Hub model with repulsive el-el couplings, $K_c^{\rm eff} < 1$ always,
which means a phase with dominant SS is impossible.
For this model the condition $K_c^{\rm eff} = 1/2$
occurs at the critical el-ph coupling value
\ba
\label{Kc12Holstein}
&& \lambda^{\rm CDW}_{\rm Pei} = {\bar U}\left[\frac{5 h({\bar U}l_0)}{2Q} + f({\bar
U}l_0)\right]^{-1} \;\; \\
\nonumber \\
\nonumber&& \;\;\;\; {\rm(} K_c^{\rm eff} = 1/2 \; {\rm transition)}.
\ea
For $\lambda_{\rm Pei} > \lambda^{\rm CDW}_{\rm Pei}$,
the SS susceptibility is never divergent.
The condition $\lambda_{\rm Pei} = \lambda^{\rm CDW}_{\rm Pei}$ determines the dashed line in
Figs. \ref{bindloss_fig_3}(c) and \ref{bindloss_fig_3}(d).

\subsection{Half-filling}

At half-filling, Eqs. (\ref{5}) and (\ref{6}) can be integrated 
analytically if one takes $g_c^0 = -g_3^0$.  For that case,
\ba
g_c(\omega) &=& -g_3(\omega) = \frac{g_c^0}{1 + g_c^0 l} , \\
\nonumber g_{+,{\rm ph}}(\omega) &=&  \left( \frac{g_{+,{\rm ph}}^0}{1 + g_{+,{\rm ph}}^0
X}\right)
\left(\frac{g_1(\omega)}{g_1^0}\right)^{3/2}\left(\frac{g_c(\omega)}{g_c^0}\right)^{-1/2},\\
\\
\nonumber g_{-,{\rm ph}}(\omega) &=&  \left( \frac{g_{-,{\rm ph}}^0}{1 + g_{-,{\rm ph}}^0
Y}\right)\left(\frac{g_1(\omega)}{g_1^0}\right)^{3/2}
\left(\frac{g_c(\omega)}{g_c^0}\right)^{3/2},
\\
\ea
where
\ba
X = \int_0^l dx \, (1 + g_1^0 x)^{-3/2} (1 + g_c^0 x)^{1/2} ,
\;\;\;\;\;\;\;\;\;\;\;\;\;\;\;\;\;\;\;\; \\
\nonumber Y = \int_0^l dx \, (1 + g_1^0 x)^{-3/2} (1 + g_c^0 x)^{-3/2}
\;\;\;\;\;\;\;\;\;\;\;\;\;\;\;\;\;\;\;\;\\
\;\; = \frac{2}{(g_1^0 - g_c^0)^2} \left[ g_1^0 + g_c^0 - \frac{g_1^0 + g_c^0 + 2 g_1^0 g_c^0
l}{\sqrt{(1 + g_1^0 l)(1 + g_c^0 l)}} \right] , \,\;\; 
\ea
and $g_1(\omega)$ is given again by Eq. (\ref{g1omega}).
Since our assumption of $g_c^0 = -g_3^0$ is satisfied for
the Hubbard model (with $V = 0$), the above results
determine the scalings of $g_{1,{\rm ph}}$ and
$g_{3,{\rm ph}}$ for the Hol-Hub and Pei-Hub models, which are
\ba
&&\label{10}\nonumber g_{1,{\rm ph}}(\omega) = g_{3,{\rm ph}}(\omega) = \left(\frac{g_{1,{\rm
ph}}^0}{1 + g_{1,{\rm ph}}^0 X^\prime}\right)\sqrt{\frac{1-Ul}{(1+Ul)^3}}\\
&&\nonumber\\
&&\;\;\;\;\;\;({\rm half{\rm-}filled \; Holstein {\rm-} Hubbard \; model}),
\ea
\ba
&&\label{11}\nonumber g_{1,{\rm ph}}(\omega) = - g_{3,{\rm ph}}(\omega) = \left(\frac{g_{1,{\rm
ph}}^0}{1 + g_{1,{\rm ph}}^0 Y^\prime}\right)\sqrt{\frac{1}{[1-(Ul)^2]^3}}\\
&&\nonumber\\
&&\;\;\;\;\;\;\;\;\;({\rm half{\rm-}filled \; Peierls {\rm-} Hubbard \; model}), 
\ea
where $X^\prime$ and $Y^\prime$ are the values of $2X$ and $2Y$, respectively, for the case
$g_1^0 = -g_c^0 = U$:
\ba
X^\prime &=& \frac{2}{U}\left[ 2 - 2 \sqrt{\frac{1 - Ul}{1 + Ul}} - \arcsin(Ul) \right]  ,\\
Y^\prime &=& \frac{2l}{\sqrt{1 - (Ul)^2}}  .
\ea

In the absence of el-el interactions, for either model, $g_{1,{\rm ph}}(\omega)$ increases in
strength as $l$ is increased
according to
$g_{1,{\rm ph}}(\omega)/g_{1,{\rm ph}}^0 = (1 - 2|g_{1,{\rm ph}}^0|l )^{-1}$.
As shown in Fig. \ref{bindloss_fig_11}(b), (d), and (f),
turning on a repulsive $U$ has the
opposite effect for the half-filled Pei-Hub model compared to the half-filled Hol-Hub model:
for the former $g_{1,{\rm ph}}(\omega)$
increases even more rapidly with increasing $l$ than before,
while for the later $g_{1,{\rm ph}}(\omega)$ increases less rapidly than before.  This is due to
the bond (site) centered
nature of the Peierls (Holstein) el-ph interaction,
and shows up as a sign difference in $g_{3,{\rm ph}}^0$ for the two models.

For $U \gg |g_{1,{\rm ph}}^0|$, the behavior of $g_{1,{\rm ph}}(\omega)/g_{1,{\rm ph}}^0$
is given by the square roots in Eqs. (\ref{10}) and (\ref{11}),
and is shown in Fig. \ref{bindloss_fig_11}(f).
In this limit, for the Hol-Hub model,
$g_{1,{\rm ph}}(\omega)$ flows to weaker values.
Since the spin gap is enhanced by a strong
$g_{1,{\rm ph}}$, and
the charge gap is enhanced by a strong
$g_{3,{\rm ph}}$, we see that
the off-diagonal phonon mechanism in the Pei-Hub model is more effective in enhancing both the
charge
and spin gap compared to the diagonal mechanism in the Hol-Hub model.

At half-filling,
the transition line between CDW and SDW phases,
called the Mott-Peierls transition,
is determined by $g_1^{\rm tot}(\omega_0) = 0$.
The critical el-ph couplings that define this
phase boundary are then
\ba
&&\label{13} \;\;\;\;\;\;\;\;\;\;\;\;\;\;\;\; \lambda^{\rm GAP}_{\rm Pei} = \frac{{\bar U}\sqrt{1
- ({\bar U}l_0)^2}}{(1 - {\bar U}l_0)^{-1} + 2 {\bar U}l_0} \\
\nonumber\\
\nonumber&&{\rm (Peierls{\rm -}Hubbard\; model{\rm :} \; \: Mott{\rm -}Peierls \; transition)} ,
\ea
\ba
&&\label{12} \;\;\;\lambda^{\rm GAP}_{\rm Hol} = {\bar U}\left[4 - 3 \sqrt{\frac{1 - {\bar
U}l_0}{1 + {\bar U}l_0}} -2 \, \arcsin({\bar U}l_0)\right]^{-1} \\
\nonumber\\
\nonumber&&{\rm (Holstein{\rm -}Hubbard\; model{\rm :} \; \: Mott{\rm -}Peierls \; transition)} 
.
\ea
For $\lambda_{\rm Pei} > \lambda^{\rm GAP}_{\rm Pei}$ in the half-filled Pei-Hub model, or
$\lambda_{\rm Hol} > \lambda^{\rm GAP}_{\rm Hol}$ in the half-filled Hol-Hub model,
the ground state is a spin-gapped, ordered CDW.
We plot the transition line given by $\lambda_{\rm Pei} = \lambda^{\rm GAP}_{\rm Pei}$ in Figs.
\ref{bindloss_fig_7}(b)
and \ref{bindloss_fig_8}, and the line determined by $\lambda_{\rm Hol} = \lambda^{\rm GAP}_{\rm
Hol}$ in Figs. \ref{bindloss_fig_7}(a) and
\ref{bindloss_fig_10}.   

\subsection{Near half-filling}

Using the two-step RG technique,
we have derived phase boundaries for the strongly incommensurate case $\mu \sim E_F$,
as well as the half-filled case $\mu = 0$.
For the more general case $0 < \mu < E_F$, in other words
at filling near but not equal to half-filling,
a three-step RG method is necessary.
The three distinct crossover scales are the high-energy $E_F$, low-energy scale $\omega_0$, and chemical potential $\mu$.
As before, retarded interactions only renormalize when integrating
out states between $E_F$ and $\omega_0$.  However, now,
$g_3$ and $g_{3, {\rm ph}}$ only play a role when integrating
out states at higher energies than $\mu$ (and
if $\mu < \omega_0$, for states between $\mu$ and $\omega_0$, only $g_3$ plays a role).

\subsubsection{Doping dependence of the phase boundaries}

We now employ the three-step RG technique to derive the 
doping dependence of the phase boundaries for $V = 0$.
First consider the case $\omega_0 < \mu < E_F$.  We begin by integrating out degrees of freedom
between $E_F$ and $\mu = \omega_0(E_F/\omega_0)^\delta$,
resulting in an effective $g_{1,{\rm ph}}$ of
\be 
g_{1,{\rm ph}}(\mu) = - \left(\frac{\pi v_F \lambda_{\rm Hol}}{1 - \lambda_{\rm Hol} \tilde X /
\bar U}\right)\sqrt{\frac{1-c{\bar U} l_0}{(1+c{\bar U} l_0)^3}}
\ee
or
\be
g_{1,{\rm ph}}(\mu) = -\left(\frac{\pi v_F \lambda_{\rm Pei}}{1 - \lambda_{\rm Pei} \tilde Y /
\bar U}\right)\sqrt{\frac{1}{[1-(c{\bar U} l_0)^2]^3}} ,
\ee
for the Hol-Hub and Pei-Hub models, respectively, with 
\ba
c &=& 1 - \delta, \\
\tilde X &=& 2\left[2 - 2 \sqrt{\frac{1 - c{\bar U} l_0}{1 + c{\bar U} l_0}} - \arcsin(c{\bar U}
l_0)\right],\\
\tilde Y &=& \frac{2c \bar U l_0}{\sqrt{1 - (c{\bar U} l_0)^2}}.
\ea 
Next, $g_{1,{\rm ph}}(\mu)$ is used as the initial value to integrate
from $\mu$ to $\omega_0$,
employing the RG flow equations without $g_3$ and $g_{3,{\rm ph}}$,
resulting in
\be
\label{g1ph3step}
g_{1,{\rm ph}}(\omega_0) = \left(\frac{g_{1,{\rm ph}}(\mu)}{1 + g_{1,{\rm ph}}(\mu) \tilde
Z/U}\right) 
\sqrt{\frac{\exp(\delta {\bar U}l_0)}{(1 + \delta {\bar U}l_0)^3}}
\ee
for either model, where
\be
\tilde Z = \int_0^{\delta {\bar U}l_0} dx \: e^{x/2}(1 + x)^{-3/2} .
\ee
Since $g_1$ renormalizes in the same way for the half-filled and incommensurate cases,
we can just integrate from $E_F$ to $\omega_0$ in one step using Eq. (\ref{g1omega}).

Again, the condition $g_1^{\rm tot}(\omega_0) = g_1(\omega_0) + g_{1,{\rm ph}}(\omega_0) = 0$
determines the transition to a spin gap,
which leads to the critical values
\ba
\label{Pei_gen}
&&\lambda_{\rm Pei}^{\rm gap} = \frac{\bar U \sqrt{[1 - (c{\bar U}l_0)^2]^3}} {4 + \xi+ 2c{\bar
U}l_0[3 - (c{\bar U}l_0)^2]} \;\;\;\;\; \\
\nonumber\\
\nonumber&&\;\;\;\;\;\;\;{\rm (Peierls{\rm -}Hubbard\; model)} ,
\ea
\ba
\label{Hol_gen}
&&\lambda_{\rm Hol}^{\rm gap} = {\bar U}\left[ 4 + \xi \sqrt{ \frac{1 - c{\bar U}l_0} {(1 +
c{\bar U}l_0)^3} } - 2 \arcsin(c {\bar U}l_0) \right]^{-1} \;\;\; \\
\nonumber\\
\nonumber&&\;\;\;\;\;\;\;\;\;\;\;\;\;\;\;\;\;\;{\rm (Holstein{\rm -}Hubbard\; model)} ,
\ea
where we defined
\ba
\xi &=&   (1 + {\bar U}l_0)F - 4(1 + c{\bar U}l_0) + \tilde Z , \\
F &=& e^{\delta {\bar U}l_0/2}(1 + \delta {\bar U}l_0)^{-3/2}  .
\ea
The system
is spin-gapped for $\lambda_{\rm Pei} > \lambda_{\rm Pei}^{\rm gap}$
or $\lambda_{\rm Hol} > \lambda_{\rm Hol}^{\rm gap}$ for the 
Pei-Hub and Hol-Hub models, respectively.   
The phase boundaries determined by $\lambda_{\rm Pei} = \lambda_{\rm Pei}^{\rm gap}$
and $\lambda_{\rm Hol} = \lambda_{\rm Hol}^{\rm gap}$
are shown in Figs. \ref{bindloss_fig_6} - \ref{fig_1} as
thick solid lines.  For $\delta = 1$,
$\lambda_{\rm Pei}^{\rm gap} = \lambda_{\rm Hol}^{\rm gap}$,
and we recover the incommensurate LL-LEL transition [Eq. (\ref{spingapeq})] with $V = 0$.

We obtain analytic expressions for the remaining phase boundaries by
requiring that $K_c^{\rm eff}$ equals 1 or 1/2 (depending on the phase boundary), using 
\be
\label{gc3step}
g_c^{\rm tot}(\omega_0) = g_c(\mu) + g_{1,{\rm ph}}(\omega_0) - 2g_{2,{\rm ph}}^0
\ee
with $g_c(\mu) = -U/(1 - c {\bar U}l_0)$.  The results for the critical couplings,
for $\omega_0 < \mu < E_F$, are then
\ba
\label{Kc1Hol}
\lambda_{\rm Hol}^{\rm{ss}, \pm} &=& {\bar U}\left[B \pm \sqrt{B^2 - S AC/2}\right] ,\\
\label{Kc12Hol}
\lambda_{\rm Hol}^{\rm cdw} &=& {\bar U}\left[D + \sqrt{D^2 + 5 S AE/4}\right] ,\\
\label{Kc12Pei}
\lambda_{\rm Pei}^{\rm cdw} &=& {\bar U}\left[ \tilde Y + \frac{F/E + \tilde Z}{\sqrt{[1 -
(c{\bar U}l_0)^2]^3}}  \right]^{-1} ,
\ea
with the definitions
\ba
A &=& (S \tilde X + \tilde Z)^{-1}  ,\\
B &=& [(2S - F)A + C]/4 ,\\
C &=& (1-c{\bar U}l_0)^{-1} ,\\
D &=& A[S(4 - 5 E\tilde X) - 5(E\tilde Z + F)]/8 ,\\
E &=& (6/{\bar U} + 3)/5 - C  ,\\
S &=& (1 + c{\bar U}l_0)^{3/2}(1 - c{\bar U}l_0)^{-1/2}  .
\ea
For $\delta = 1$, 
Eqs. (\ref{Kc1Hol}), (\ref{Kc12Hol}), and (\ref{Kc12Pei}) reduce to Eqs.
(\ref{18}), (\ref{21}), and (\ref{Kc12Holstein}), respectively, with $V = 0$.
The conditions $\lambda_{\rm Hol} = \lambda_{\rm Hol}^{\rm{ss}, +}$ and
$\lambda_{\rm Hol} = \lambda_{\rm Hol}^{\rm{ss}, -}$
determine the thin solid line in Figs. \ref{Hol_doping} and \ref{fig_1}(a).  In Fig.
\ref{fig_1}(b), ${\bar U}$ is large enough such that 
$S AC > 2B^2$ everywhere in the plot, therefore the phase with dominant SS is not present.
The condition $\lambda_{\rm Hol} = \lambda_{\rm Hol}^{\rm cdw}$ determines the dashed line in
Figs. \ref{Hol_doping}, \ref{fig_1}(a), and \ref{fig_1}(b). The dashed line in Figs.
\ref{Pei_doping}, \ref{fig_1}(c), and \ref{fig_1}(d) are determined by $\lambda_{\rm Pei} =
\lambda_{\rm Pei}^{\rm cdw}$.

For the case $\mu < \omega_0 < E_F$, since
$g_1$ scales in the same way at all energies and does not depend on $g_3$,
the spin-gap phase boundary
is independent of $\mu$, and is given by
Eqs. (\ref{Pei_gen}) and (\ref{Hol_gen}) with $\delta = 0$, or by
Eqs. (\ref{13}) and (\ref{12}).

\subsubsection{Doping dependence of susceptibilities and isotope effects}

The doping dependence of the susceptibilities in the LEL phase (Figs. \ref{fig_2} and
\ref{fig_3}) is computed with the three-step RG method
using Eqs. (\ref{barchiSS}) and (\ref{barchiCDW}), combined with Eqs. (\ref{15}),
(\ref{computedelta}), (\ref{g1ph3step}), and (\ref{gc3step}).
 
The isotope effect on $T_c$ (Fig. \ref{fig_4}) is computed via
\be
\alpha_{T_c}  = -\frac{1}{2}\frac{\Delta T_c/T_c} {\Delta \zeta/\zeta},
\ee
where $\zeta \equiv E_F/\omega_0 = e^{l_0}$ and $\Delta \zeta \ll \zeta$.
Here, $\Delta T_c \equiv T_c^\prime - T_c$, where
$T_c^\prime$ is the transition temperature determined from Eq. (\ref{determineTc})
after changing $l_0 \rightarrow l_0^\prime = \ln({\zeta} + 
\Delta {\zeta})$, $c \rightarrow c^\prime = c \, l_0/l_0^\prime$, and
$\delta \rightarrow 1 - c^\prime$.  The changes in $c$ and $\delta$ are required so that,
when changing $\zeta$, only the energy scale $\omega_0$ is changed, and the energy scales $E_F$
and $\mu$ remain fixed.  The isotope effect on $\Delta_s$ is determined in a similar fashion.

\section{Conclusions}
\label{conclusion}

We have explored the 
influence of the el-ph interaction on the
quantum phase diagram of the most theoretically well
understood non-Fermi liquid,
the interacting 1DEG.
The backward and Umklapp scattering portions of the
el-ph interaction are strongly renormalized, often toward stronger couplings. 
Even in the presence of strong el-el repulsion,
a weak, retarded el-ph interaction is capable
of creating a spin gap and causing divergent superconducting and/or CDW susceptibilities
(true long range order is formed
when weak coupling between 1D chains is included).
The ground state is strongly dependent on the band-filling, and, especially at or near
half-filling, dependent on the microscopic model of the el-ph interaction.
Compared to higher dimensions, the zero-temperature phase
diagram is far more complex, and, away from commensurate filling,
contains a subtle competition between SDW, CDW, and superconductivity.
The fact that direct el-el interactions
strongly influence the renormalizations of the el-ph interactions adds to the richness
of the phase diagram.  In 1D, intuitive concepts that apply to higher dimensional Fermi liquids,
such as the suppression of superconductivity by repulsive interactions at weak coupling, must
sometimes be abandoned.

When the bare el-el repulsion is much stronger than the bare el-ph induced attraction, in 1D,
unlike in higher dimensions, it is not a requirement that $E_F \gg \omega_0$ for the
superconducting susceptibility to diverge. (In fact, in 1D, very large values of $E_F/\omega_0$
are harmful to superconductivity.)  Note that in the high-temperature superconductors, where
$E_F/\omega_0 \sim 5$ and the el-el repulsion is strong, it has been correctly argued that the
small value of $E_F/\omega_0$ rules out conventional phonon-mediated
superconductivity.\cite{KivelsonBook}  It is interesting to point out that the arguments there
apply only to a Fermi liquid and not the quasi-1DEG.

We now qualitatively summarize the phase diagrams, for the case of repulsive el-el interactions,
beginning with a system that is far from half-filling.  In this case, the charge sector is
gapless. For either the Hol-Hub or Pei-Hub model, the spin-gapped LEL phase is favored by small
$U$, large $V$, large $\lambda$, and large $E_F/\omega_0$.  The LL phase is favored by large $U$,
small $V$, small $\lambda$, and small $E_F/\omega_0$.  For the Hol-Hub model, a dominant
superconducting fluctuation is favored by small $U$, small $V$, moderate $\lambda_{\rm Hol}$, and
small $E_F/\omega_0$.  For either model, the phase with dominant $2 k_F$ CDW and subdominant
superconductivity is favored by moderate $E_F/\omega_0$ and moderate $\lambda$ (the dependence on
$U$ and $V$ is more subtle--see the phase diagrams).  For either model, the phase with dominant
$2 k_F$ CDW and subdominant $4 k_F$ CDW is favored by large $V$, large $\lambda$, and large
$E_F/\omega_0$ (see the diagrams for the subtle dependence on $U$).

Moving the incommensurate system toward half-filling increases the stability of the LEL phase
relative to the LL phase in the Pei-Hub model, but decreases the stability of the LEL phase
relative to the LL phase in the Hol-Hub model.  In the Hol-Hub model, moving toward half-filling
suppresses the phase with dominant superconductivity.  For both models, moving toward
half-filling decreases the stability of the phase with subdominant SS and increases the
stability of the phase with subdominant $4 k_F$ CDW, but this effect is more pronounced in the
Pei-Hub model.

In the Hol-Hub model at half-filling, a spin-gapped CDW phase is favored by small $U$ and large
$\lambda_{\rm Hol}$.  The SDW phase with no spin gap is favored by large $U$ and small
$\lambda_{\rm Hol}$.  The half-filled Hol-Hub phase diagram is weakly dependent on $E_F/\omega_0$
compared to the Pei-Hub phase diagram.  In the half-filled Pei-Hub model, the spin-gapped CDW
phase is favored by large $\lambda_{\rm Pei}$, large $E_F/\omega_0$, large $V$, and any $U$ other
than moderate values.  The SDW phase is favored by small $\lambda_{\rm Pei}$, small
$E_F/\omega_0$, small $V$, and moderate $U$.  Both models are charge-gapped at half-filling.   

We have studied the strong doping dependencies of the phonon-induced spin gap and various
susceptibilities in the Luther-Emery liquid phase.  The spin gap and charge density wave
susceptibilities decrease monotonically as the system is doped away from half-filling.  However,
the superconducting susceptibility, and therefore $T_c$ in a quasi-1D system with fluctuating
chains, can vary nonmonotonically with doping and exhibit a maximum at some ``optimal'' doping.

Partially motivated by the unconventional doping-dependent isotope effects observed in the
cuprate high-temperature superconductors, we have computed isotope effects in the quasi-1DEG
coupled to phonons, since it is perhaps the most easily studied unconventional phonon-mediated
superconductor.  The calculated isotope effects bear a qualitative resemblance to those observed
in the cuprates, as summarized in Section \ref{introduction}.
\begin{acknowledgments}

I would like to thank S. E. Brown, E. Fradkin, and especially S. Kivelson for helpful
conversations.
This work was supported by the Department of Energy Contract 
No. DE-FG03-00ER45798.

\end{acknowledgments}

\end{document}